\documentclass[3p,authoryear,preprint,12pt]{elsarticle}
\makeatletter\if@twocolumn\PassOptionsToPackage{switch}{lineno}\else\fi\makeatother

\usepackage{tabulary,xcolor}
\usepackage{tablefootnote}
\usepackage{amsfonts,amsmath,amssymb}
\usepackage[T1]{fontenc}
\makeatletter
\let\save@ps@pprintTitle\ps@pprintTitle
\def\ps@pprintTitle{\save@ps@pprintTitle\gdef\@oddfoot{\footnotesize\itshape \null\hfill\today}}
\def\hlinewd#1{%
  \noalign{\ifnum0=`}\fi\hrule \@height #1%
  \futurelet\reserved@a\@xhline}

\AtBeginDocument{\ifNAT@numbers \biboptions{sort&compress}\fi}
\makeatother

\usepackage{caption}
\usepackage{ifluatex}
\ifluatex
\usepackage{fontspec}
\defaultfontfeatures{Ligatures=TeX}
\usepackage[]{unicode-math}
\unimathsetup{math-style=TeX}
\else 
\usepackage[utf8]{inputenc}
\fi 
\ifluatex\else\usepackage{stmaryrd}\fi

\usepackage{url,multirow,morefloats,floatflt,cancel,tfrupee}
\makeatletter
\AtBeginDocument{\@ifpackageloaded{textcomp}{}{\usepackage{textcomp}}}
\makeatother
\usepackage{colortbl}
\usepackage{graphicx}
\usepackage{xcolor}
\usepackage{pifont}
\usepackage[nointegrals]{wasysym}
\usepackage{natbib,hyperref}
\hypersetup{colorlinks,allcolors=blue,pdfauthor=author}
\usepackage{xcolor}
\usepackage{float}
\usepackage{lipsum}
\usepackage{listings}
\usepackage{subcaption}
\usepackage{stfloats}
\makeatletter
\newcommand{\rcolor}[1]{{#1}}
\def\mcWidth#1{\csname TY@F#1\endcsname+\tabcolsep}

\def\cAlignHack{\rightskip\@flushglue\leftskip\@flushglue\parindent\z@\parfillskip\z@skip}
\def\rAlignHack{\rightskip\z@skip\leftskip\@flushglue \parindent\z@\parfillskip\z@skip}

\@ifundefined{etal}{}{}

\if@twocolumn\usepackage{dblfloatfix}\fi
\usepackage{ifxetex}
\ifxetex\else\if@twocolumn\@ifpackageloaded{stfloats}{}{\usepackage{dblfloatfix}}\fi\fi

\AtBeginDocument{
\expandafter\ifx\csname eqalign\endcsname\relax
\def\eqalign#1{\null\vcenter{\def\\{\cr}\openup\jot\m@th
  \ialign{\strut$\displaystyle{##}$\hfil&$\displaystyle{{}##}$\hfil
      \crcr#1\crcr}}\,}
\fi
}

\AtBeginDocument{%
  \@ifpackageloaded{endfloat}%
   {\renewcommand\efloat@iwrite[1]{\immediate\expandafter\protected@write\csname efloat@post#1\endcsname{}}}{\newif\ifefloat@tables}%
}%

\def\BreakURLText#1{\@tfor\brk@tempa:=#1\do{\brk@tempa\hskip0pt}}
\let\lt=<
\let\gt=>
\def\processVert{\ifmmode|\else\textbar\fi}

\@ifundefined{subparagraph}{
\def\subparagraph{\@startsection{paragraph}{5}{2\parindent}{0ex plus 0.1ex minus 0.1ex}%
{0ex}{\normalfont\small\itshape}}%
}{}

\newcommand\role[1]{\unskip}
\newcommand\aucollab[1]{\unskip}
  
\@ifundefined{tsGraphicsScaleX}{\gdef\tsGraphicsScaleX{1}}{}
\@ifundefined{tsGraphicsScaleY}{\gdef\tsGraphicsScaleY{.9}}{}
\def\checkGraphicsWidth{\ifdim\Gin@nat@width>\linewidth
	\tsGraphicsScaleX\linewidth\else\Gin@nat@width\fi}

\def\checkGraphicsHeight{\ifdim\Gin@nat@height>.9\textheight
	\tsGraphicsScaleY\textheight\else\Gin@nat@height\fi}

\def\fixFloatSize#1{}
\let\ts@includegraphics\includegraphics

\def\inlinegraphic[#1]#2{{\edef\@tempa{#1}\edef\baseline@shift{\ifx\@tempa\@empty0\else#1\fi}\edef\tempZ{\the\numexpr(\numexpr(\baseline@shift*\f@size/100))}\protect\raisebox{\tempZ pt}{\ts@includegraphics{#2}}}}

\AtBeginDocument{\def\includegraphics{\@ifnextchar[{\ts@includegraphics}{\ts@includegraphics[width=\checkGraphicsWidth,height=\checkGraphicsHeight,keepaspectratio]}}}

\DeclareMathAlphabet{\mathpzc}{OT1}{pzc}{m}{it}

\def\URL#1#2{\@ifundefined{href}{#2}{\href{#1}{#2}}}

\def\UrlOrds{\do\*\do\-\do\~\do\'\do\"\do\-}%
\g@addto@macro{\UrlBreaks}{\UrlOrds}

\edef\fntEncoding{\f@encoding}

\makeatother

\newif\ifmultipleabstract\multipleabstractfalse%
\begin{document}
\pagenumbering{arabic}


\begin{frontmatter}
    \title{
  Physical properties of the slow-rotating near-Earth asteroid (2059) Baboquivari from one apparition
}
    
\author[ubtakdeniz,tubitaktug]{Orhan Erece\corref{ereceemail}}
\ead{orhanerece@gmail.com }\cortext[ereceemail]{Corresponding author.}
\author[kfu,ant]{Irek M. Khamitov}
\author[ubtakdeniz]{Murat Kaplan}
\author[ubtakdeniz,tubitaktug]{Yucel Kilic}
\author[kasi]{Hee-Jae Lee}
\author[kasi]{Myung-Jin Kim}
\author[kfu,ant]{Ilfan F. Bikmaev}
\author[kfu,ant]{Rustem I. Gumerov}
\author[kfu,ant]{Eldar N. Irtuganov}

\address[ubtakdeniz]{Department of Space Sciences and Technologies\unskip, 
    Akdeniz University\unskip, Campus\unskip, Antalya\unskip, 07058\unskip, Turkey}
  	
\address[tubitaktug]{
    T{\"{U}}B{\.{I}}TAK National Observatory\unskip, Akdeniz University Campus\unskip, Antalya\unskip, 07058\unskip, Turkey}

\address[kfu]{Kazan Federal University, Kremlevskaya Str., 18, Kazan, 420008, Russia}
\address[ant]{Academy of Sciences of Tatarstan Republic, Bauman Str., 20, Kazan, 420111, Russia}

\address[kasi]{Korea Astronomy and Space Science Institute, 776, Daedeokdae-ro, Yuseong-gu, Daejeon 34055, Korea}

\begin{abstract}
In this study, we carried out photometric, spectroscopic, and for the first time, polarimetric observations of the Amor-type near-Earth asteroid (2059) Baboquivari. Our findings represent the first reliable determination of Baboquivari's physical properties. We used data from a 1m-class telescope (T100) along with ALCDEF data for photometric analyses and a 1.5-m-class telescope (RTT150) for polarimetric, spectroscopic, and additional photometric observations. We obtained the synodic rotation period of Baboquivari as $129.93 \pm 2.31$ hours and the standard phase function parameters $H$ and $G$ as $16.05 \pm 0.05$, $0.22 \pm 0.02$, respectively. Our colour index (V-R) measurement of $0.45 \pm 0.02$ is consistent with spectroscopic observations, indicating an S (or sub-S) spectral type. Using the polarimetric and spectroscopic data, we found that the geometric albedo is $0.15 \pm 0.03$, and the spectral type is Sq. Based on the estimated albedo and absolute magnitude, Baboquivari has an effective diameter of $2.12 \pm 0.21$ km. Due to the scattered data in the light curve, its slow rotation and location among the NEAs suggest that Baboquivari may be a non-principal axis (NPA) rotator.
\end{abstract}
  
\end{frontmatter}

\section{Introduction}
Studies on determining the physical properties of near-Earth asteroids (NEAs) are important for a better understanding of their nature and have become more critical with the increase of in-situ asteroid missions. While determining the target asteroid for a mission, features such as shape, size, and composition are definitely taken into account. Various ground-based observation techniques, such as photometry, polarimetry, and spectroscopy, should be used to reveal these features. The observational data obtained by all three methods give different features about the asteroid. The interpretation and combination of these three observation techniques provide essential and more accurate elements.

The physical properties of asteroids are sensitively related to each other. For example, a well-known relation between the diameter ($D_{eff}$), absolute magnitude ($H$), and geometric albedo ($p_{V}$) (e.g., \cite{FowlerChillemi,muinonen2010}) have a high-frequency of use. When determining the diameter of an asteroid, estimating the absolute magnitude, which is one of the required values, is possible from photometric observations; however, observational (polarimetric, radiometric, etc.) determination of albedo is essential for an accurate result. Otherwise, only conjecturing can be made using one type of data set. It should be noted that accurate estimations of albedos and diameters are critical to determining the size and albedo distributions of the asteroid population \citep{cellino2009}. In addition to photometry and polarimetry, spectroscopic measurements give direct information about the object's spectral type. Besides, radiometric and multi-chord stellar occultation techniques are crucial for more accurate size and shape determination \citep{ortiz2020}.

In this study, we carried out photometric, polarimetric, and spectroscopic observations for the Amor type NEA (2059) Baboquivari 1963 UA (hereinafter Baboquivari) was first discovered on October 16, 1963 at the Goethe Link Observatory, Indiana University. Being a slow-rotating NEA, having a relatively high eccentric orbit, and only appearing for long periods of time, our interest in this asteroid has increased. Close approaches of Baboquivari in the recent past/future to Earth can be seen in Table \ref{tab:2059_observability}\footnote{\url{https://ssd.jpl.nasa.gov/horizons/app.html}}. Those days are observationally promising for most ground-based telescopes. Here, our observations cover the days near the close approach on October 11, 2019. Due to the observational difficulties, very detailed studies have not been carried out on Baboquivari. \rcolor{The first calculations were made by  \citet{veres2015} who obtained $H = 16.45^{m} \pm 0.59$ and $G = 0.80 \pm 0.41$ using the $HG$ model \citep{bowell1989} parameters while $HG_{12}$ model \citep{muinonen2010} parameters $H = 16.30^{m} \pm 0.32$ and $G_{12}= -0.18 \pm 0.23$} from Pan-STARRS1 telescope all-sky survey mission with relatively high uncertainty. \citet{warnerstephens2020} found Baboquivari's period 129.43 hours and reported the object as a possible member of the very wide binary asteroid class or a tumbler with a low amplitude component based on its long period. In \cite{masiero2021}, a couple hundred NEOs' diameters and albedos were computed from the Near-Earth Object Wide-field Infrared Survey Explorer (NEOWISE) observations. Using thermal modeling technique and assuming $G$ = 0.15, they found Baboquivari's \rcolor{$H = 15.90^{m}$, $D_{eff} = 1.84 \pm 0.79$ km and $p_v = 0.179^{+0.188}_{-0.092}$}. However, those results are also highly uncertain. \rcolor{\cite{sonka2021}, on the other hand, viewed Baboquivari from a different perspective and stated the object as a suitable target for space missions due to its low $\Delta v$ velocity. They also found two possible rotational periods of Baboquivari as $133.3222\pm0.1$ hours with an amplitude of 0.48 and $152.1 \pm 0.1$ hours with an amplitude of 0.63.} Besides, unpublished IR spectral data of Baboquivari are available in MIT-Hawaii Near-Earth Object Spectroscopic Survey (MITHNEOS) \rcolor{database\footnote{\url{http://smass.mit.edu/minus.html}} \citep{binzelmithneos}}. Using those data, the online implementation of the Bus-DeMeo taxonomy\footnote{\url{http://smass.mit.edu/cgi-bin/busdemeoclass-cgi}} \citep{demeo2009} classifies the most probable spectral type of Baboquivari as S, Sq, or Q type.

In Section \ref{sec:observations}, we explain our ground-based observations as to how we acquired images of Baboquivari and combined them with archival data such as the Asteroid Lightcurve Data Exchange Format (ALCDEF) \citep{warner2011} and MITHNEOS databases. Section \ref{sec:analysis} details a photometry reduction and analysis routine using Gaia Data Release 3 (DR3) data, as well as polarimetry and spectroscopy. In section 4, we give the physical properties of Baboquivari based on our analysis and its place in NEA distribution as well as future plans, missing parts, and other contributions to this field of study.
 
\begin{table*}
\caption{Close approaches of Baboquivari in the recent past/future to Earth}
\label{tab:2059_observability}
\centering 
\begin{tabular}{l c c c c c }
\hline 
\text{Date} & \text{Nominal distance (au)} & \rcolor{\text{Apparent mag.(V)}} \\ 
\hline
1963-10-20 & 0.25052 & 13.9 \\
1976-10-25 & 0.43060 & 16.3 \\
2019-10-11 & 0.47426 & 16.5 \\
2032-10-24 & 0.34652 & 15.6 \\
2075-10-20 & 0.31429 & 15.3 \\
2088-10-26 & 0.39975 & 16.1 \\

\hline 
\end{tabular}
\end{table*}

\section{Observations}\label{sec:observations}
Ground-based observations of Baboquivari were conducted using T100 and RTT150 telescopes located at TÜBİTAK National Observatory, Antalya, Turkey, at an altitude of 2500 meters (MPC code A84). We have taken 425 photometric, 5 polarimetric, and 3 spectroscopic images in total. \rcolor{Nightly seeing values were satisfying and not more than $1.6"$ for each observation night (see Table \ref{tab:t100obs-details}).} Together with our own photometric observations, the Asteroid Lightcurve Data Exchange Format
(ALCDEF; \cite{warner2011}) database has been used to extend the photometric data of Baboquivari. In the ALCDEF database, there are 827 observational data taken between July 21, 2019 and August 23, 2019, while Gaia DR3 has 134 observations between September 11, 2015 and March 25, 2016. Unfortunately, Gaia DR3 data seems not usable due to the probable strong effect of aspect changes as well as unknown rotational and physical properties.
\rcolor{Our visible spectra, obtained in the 4000\AA\ - 7500\AA\ range have been acquired at the RTT150 telescope. These observations were combined with the IR spectra made available by the MITHNEOS to complete the wavelength coverage (see Section \ref{subsec:spectroscopy}). Spectra of Baboquivari in the 6500\AA\ - 25000\AA\ range were taken on September 26, 2019 using the SpeX instrument \citep{rayner2003}. The combined spectrum of Baboquivari consists of 18 spectra in total, with 119.5 seconds of exposure each.}
\subsection{T100 Observations}\label{sec:t100}
T100 telescope is a 1-m telescope equipped with a cryo-cooler ($-95 ^{\circ} C$) SI 1100 CCD camera with $4096 \times 4096$ pixels and \rcolor{0.31$''$} angular scale per pixel. The telescope is dedicated to photometric observations only, and CCD makes it possible to reach the effective field of view (FoV) as $21' \times 21'$. \rcolor{We conducted photometric observations for 5 different nights listed in Table \ref{tab:t100obs-details}, mostly with the $V$-band (Johnson-Cousins system) and a few with the R-band to determine the V-R colour. We used an appropriate exposure time determined by considering the asteroid's sky motion rate and nightly seeing conditions in order to remain the asteroid as a point source.} We employed $2 \times 2$ pixel-binning to images and corrected each of them by subtracting bias and dividing flat in the standard way. Since the dark current and gain of SI 1100 CCD camera\footnote{\url{https://tug.tubitak.gov.tr/en/teleskoplar/t100-telescope}} are 0.0001 $e^{-}$/$pixel/s$ and 0.57 $e^{-}/ADU$ respectively, dark correction is not required up to 5,700 seconds, so we did not apply the dark correction. 

\subsection{RTT150 Observations}\label{sec:rtt150}

1.5-m Russian-Turkish telescope (RTT150) \citep{AslanRTT150} has two focal planes, Cassegrain and COUDE, respectively. Using TFOSC\footnote{\url{https://tug.tubitak.gov.tr/en/teleskoplar/rtt150-telescope-0}} (TUG Faint Object Spectrograph and Camera) instrument on Cassegrain plane and later integrated double wedged Wollaston type polarimeter by \cite{helhel2015} called TFOSC-WP, we carried out photometric, spectroscopic, and polarimetric observations listed in Table \ref{tab:t100obs-details}. The CCD camera, which is an Andor iKon-L 936\footnote{\url{https://andor.oxinst.com/products/ikon-xl-and-ikon-large-ccd-series/ikon-l-936}} with a BEX2-DD-9ZQ light detector ($-80 ^{\circ} C$) attached to TFOSC has an $11.3' \times 11.3'$ field of view, $2048 \times 2048$ pixels, and its angular scale is \rcolor{0.33$''$} per pixel in direct imaging mode. \rcolor{For photometric observations, the same technique was used as on T100 telescope, while non-sidereal tracking was used for spectroscopic and polarimetric observations.}

\begin{table*}
\caption{Ground-based observations \rcolor{and archival data} details of Baboquivari used in this work. \text{$\lambda$} is the ecliptic longitude and \text{$\beta$} is the ecliptic latitude as viewed by observer. \text{$\alpha$} is phase angle, \text{$r$} and \text{$\Delta$} are heliocentric distance and distance from observer respectively. \rcolor{Seeing is the median value and Mag.error is the mean ($\mu$) value of magnitude errors of Baboquivari derived from all the images collected throughout the night for T100 and RTT150. Seeing value for MITHNEOS is the measurement of the relevant date from the Maunakea MASS instrument and Mag.error for ALCDEF is the mean value of all 827 data.}}
\label{tab:t100obs-details}
\centering 
\resizebox{\linewidth}{!}{\begin{tabular}{c c c c c c c c c c c}
\hline 
\text{Date} & \text{Source} & \text{$\lambda$} ($^\circ$) & \text{$\beta$} ($^\circ$) & \text{$\alpha$} ($^\circ$) & \text{$r$} (AU) & \text{$\Delta$} (AU) & \text{$\#$ of obs.} & \rcolor{\text{Seeing} (")} & \rcolor{\text{Mag.error ($\mu$)}} & \text{Obs.type}\\ 
\hline
2019-08-29 & T100 & 322.0 & 23.6 & 17.7 &1.52 & 0.55 & 12 & \rcolor{1.4}  & \rcolor{0.017} & Phot\\
2019-08-31 & T100 & 321.6 & 23.4 & 18.5 & 1.51 & 0.54 & 123 & \rcolor{1.6}  & \rcolor{0.018} & Phot\\
2019-09-18 & T100 & 318.5 & 19.3 & 27.9 & 1.41 & 0.49 & 74 & \rcolor{1.4}  & \rcolor{0.026} & Phot\\
2019-09-22 & T100 & 318.4 & 18.0 & 30.1 & 1.39 & 0.48 & 56 & \rcolor{1.5}  & \rcolor{0.017} & Phot\\
2019-09-28 & T100 & 318.5 & 15.9 & 33.5 & 1.36 & 0.47 & 94 & \rcolor{1.5}  & \rcolor{0.011} & Phot\\
2019-09-27 & RTT150 & 318.5 & 16.2 & 32.9 & 1.27 & 0.48 & 60 & \rcolor{1.0} & \rcolor{0.015} & Phot\\
2019-10-20 & RTT150 & 324.3 & 6.7 & 43.6 & 1.38 & 0.47 & 6 & \rcolor{1.5} & \rcolor{0.016} & Phot\\
2019-10-20 & RTT150 & 324.3 & 6.7 & 43.6 & 1.38	& 0.47 & 5 & \rcolor{1.5} & - & Pol\\
2019-09-23 & RTT150 & 318.3 & 17.7 & 30.8 & 1.39 & 0.48 & 3 & \rcolor{1.4} & - & Spe\\
\rcolor{2019-07-21/-08-23} & \rcolor{ALCDEF} & \rcolor{328.3-323.6} & \rcolor{21.9-24.2} & \rcolor{20.0-15.8} & \rcolor{1.77-1.56} & \rcolor{0.85-0.58} & \rcolor{827} & - & \rcolor{0.025} & \rcolor{Phot}\\
\rcolor{2019-09-26} & \rcolor{MITHNEOS} & \rcolor{318.4} & \rcolor{16.8} & \rcolor{32.12} & \rcolor{1.37} & \rcolor{0.48} & \rcolor{1} & \rcolor{0.19} & - & \rcolor{Spe}\\
\hline 
\end{tabular}}
\raggedright
\textbf{Phot:} Photometry, \textbf{Pol:} Polarimetry, \textbf{Spe:} Spectroscopy\\
\end{table*}

\section{Data Reduction and Analysis}\label{sec:analysis}

\subsection{Photometry}\label{sec:photometry}
We developed a pipeline using $Python$ language and primarily based on the $Astropy$-affiliated package $Photutils$ \citep{bradley2020, astropy2013, astropy2018} for the photometry task. To validate it, we compared our photometric results with well-known $IRAF$'s $phot$ task \citep{iraf1986, irafstetson1987}. Our pipeline is capable of performing photometry on both moving objects and stars. In our algorithm, we first make astrometric calibration of the field via $astrometry.net$ \citep{lang2010} and then find all sources' physical positions ($x, y$) and celestial coordinates ($\alpha$, $\delta$) in the image using $SExtractor$ software \citep{sextractor}. Using the VizieR module in $astroquery$ package \citep{ginsburg2019} we match those sources with the Gaia DR3 catalogue and obtain existing Gaia sources in the image. We will use those cross-matched Gaia sources later to obtain zero-point magnitude as described in Section \ref{subsec:magtrans}. To generate ephemeris and find the asteroid of interest, $JPL$ $Horizons$ module in $astroquery$ package is used. Lastly, the suitable \rcolor{circular} aperture is applied for estimating the instrumental magnitude of each source as well as the asteroid. In order to transform the instrumental magnitude to standard magnitude, the calculated zero-point magnitude is added to that of the instrumental magnitude. 

\subsubsection{Zero-point magnitude calculation using Gaia DR3}\label{subsec:magtrans}
$Gaia\ DR3$ was released on 13 June 2022. It contains nearly 2 billion sources with $G$-band photometry and more than 1.5 billion sources with mean $G_{RP}$-band photometry and mean $G_{BP}$-band photometry \citep{gaia2021}. Since Gaia sources are now expected to be present in any images taken from telescopes, zero-point magnitude calculation for standard magnitude transformation can be made using these stars. To do this, we first transformed $G$-band magnitude to our interested filter Johnson-Cousins $V$-magnitude using Eq. \eqref{eq:magtransformation}. $a_{i}$ constants can be found in \cite{gaia_mag_transform2022}'s work.

\begin{equation}
\label{eq:magtransformation}
V = G + a_i(G_{RP}-G_{BP})^i
\end{equation}

Therefore, the $V$-magnitude of a star which is in the Gaia DR3 catalogue can be calculated using $G$-band magnitude and $G_{RP}$-$G_{BP}$ colour of that star. Especially in T100 observations, we find in our $21' \times 21'$  images not less than 50 sources, which are presented in the Gaia catalogue as well. After transforming their $G$-band magnitude to standard $V$-magnitude, we can calculate the zero-point magnitude for each individual image. To obtain the zero-point magnitude, we take the differences between all individual instrumental magnitudes and $G$-band magnitudes and then find a relation with the distance to the centre of the image  \rcolor{for T100 telescope despite all the pre-calibrations, e.g., flat field correction (see \ref{sec:appendixA}).} By applying a linear fit to those points and Random sample consensus (RANSAC) algorithm \citep{fischler1981} to remove outliers, we now have a specific zero-point magnitude for a specific position on the CCD surface. Although it is not the subject of this study, we can clearly see that the magnitude of the source depends on its position on the CCD surface. Thus, we can derive the standard $V$-magnitudes of the target sources according to their physical positions $\it{x,y}$ on the image. This method allows us to compare images affected by different atmospheric conditions and construct the light curve. It should be noted that the linear relation of zero-point magnitude and x, y positions can be applied to any sources in the image for reducing instrumental $V$-magnitude to standard $V$-magnitude.

\subsubsection{Rotation period and photometric phase curve}\label{subsec:hg}
\begin{figure}
\centering
\includegraphics[keepaspectratio=true,scale=0.7]{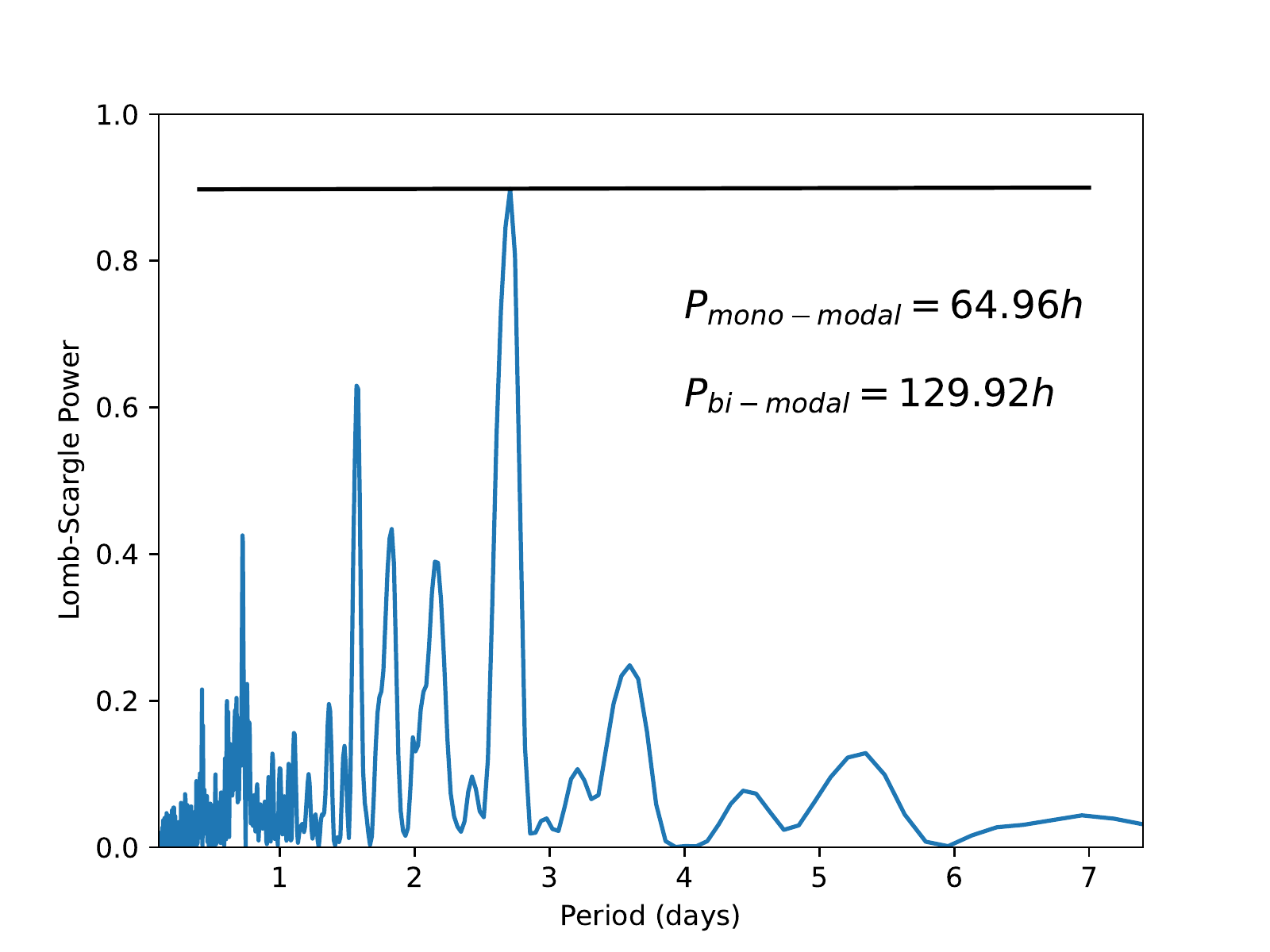}
\caption{Lomb-Scargle frequency against period in days. The graph gives a mono-modal solution as the highest frequency, yet we prefer a bi-modal solution for the period corresponding to 129.92 hours.}
\label{fig:lombscargle}
\end{figure}

\begin{figure}
\centering
\includegraphics[keepaspectratio=true,scale=0.7]{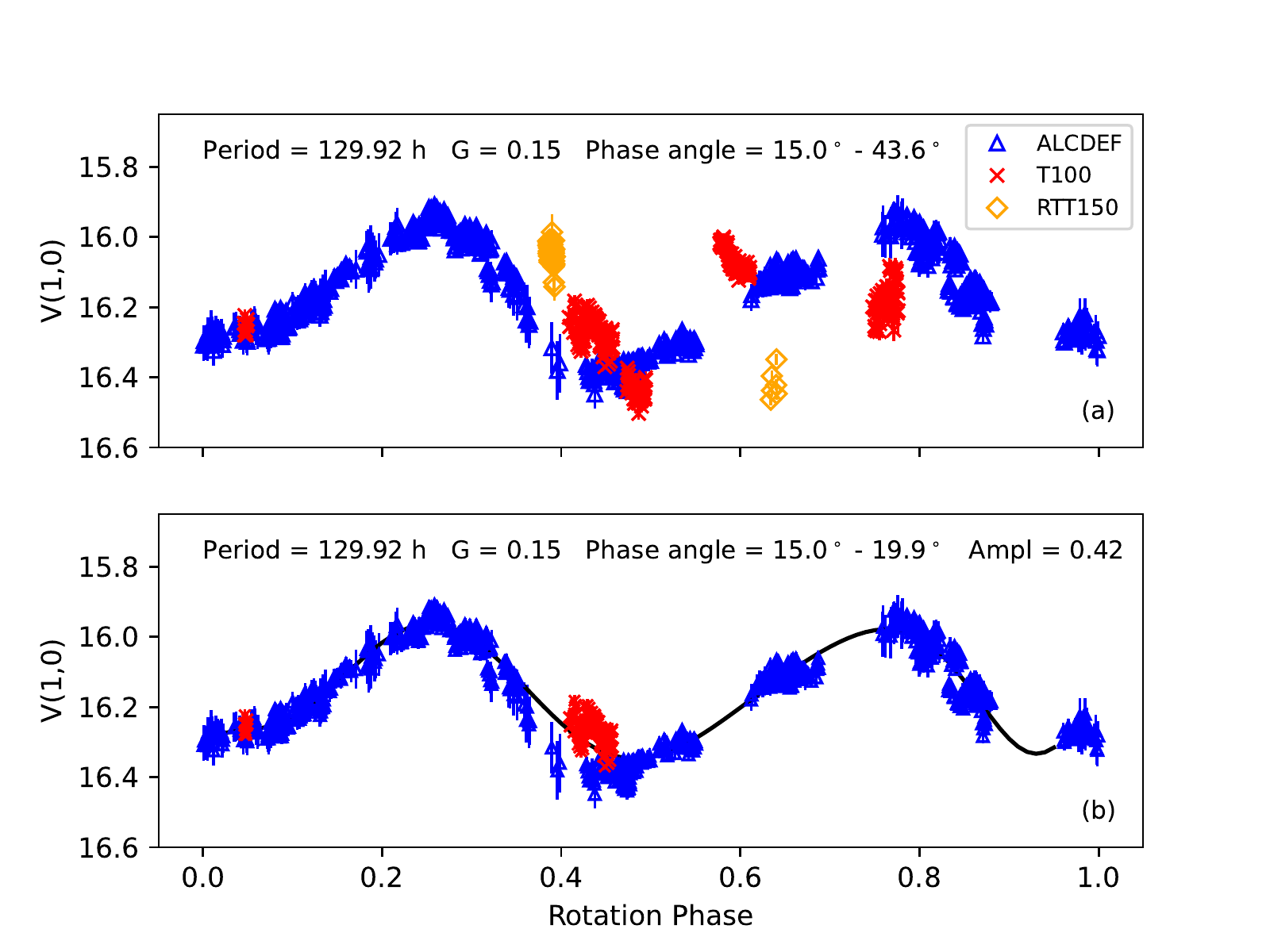}
\caption{In graph (a), all observational data were used to generate the composite light curve. As can be clearly seen from the graph, \rcolor{a fit could not be applied due to the scatter of the data associated with} NPA rotation, cometary activity, probability of being binary or aspect changes. Phasing of those data was done using the period obtained from graph (b). The composite light curve in graph (b) was generated using only ALCDEF and T100 data in \rcolor{similar} phase angles. Because we found repetitive periods, we used only those data in finding periods with the Lomb-Scargle method.}
\label{fig:lightcurve}
\end{figure}

\begin{figure}[htbp]
\centering
\includegraphics[keepaspectratio=true,scale=0.5]{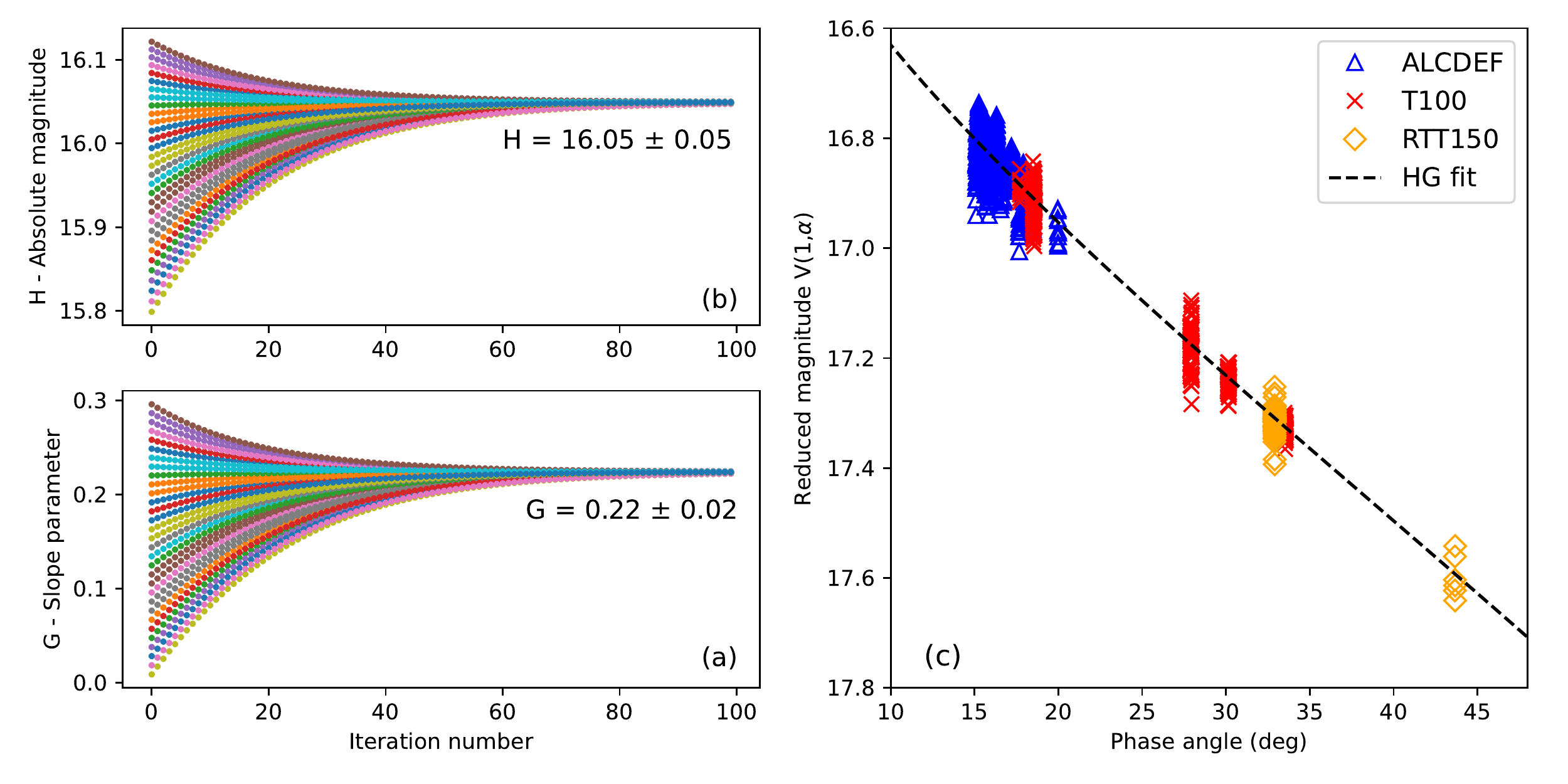}
\caption{\textbf{(a)} shows different initial $G$ values converging to a constant value while \textbf{(b)} shows the same converging situation for $H$. \textbf{(c)} Generated two-parameter $HG$ phase curve of (2059) Baboquivari using estimated values from (a) and (b).}
\label{fig:HG_phasecurveanditeration}
\end{figure}

\rcolor{Assuming that the asteroid is at 1 AU from both the Sun and Earth at 0\textdegree\ solar phase angle (angle between the Sun and Earth as seen from the asteroid), its mean apparent magnitude in the Johnson-V band over a full rotation period is called absolute magnitude ($H$). The calculation of $H$ is possible if the relation between reduced magnitude (magnitude with distance effect eliminated) and solar phase angle is well known. The relation of this so-called two-parameter $HG$-magnitude phase function, which was used in this work, was described by \cite{bowell1989}. 
For the very first estimation of asteroid properties, when observational data are lacking, the $G$ parameter (also called the slope parameter) is considered equal to 0.15.} In order to calculate the reliable rotation period of Baboquivari, we used only ALCDEF and T100 data, only which are in \rcolor{similar phase angles (between the range $15^{\circ}-20^{\circ}$)}, to avoid being affected by aspect changes. \rcolor{In our algorithm,} we first transformed apparent magnitudes into reduced magnitudes. 
Then, initially assuming $G = 0.15$, we obtained V(1,0) values from V(1,$\alpha$) magnitudes. Each V(1,0) value is now corresponding to different phases of rotation of absolute magnitudes. To find possible rotational periodicities, we used the \textit{Lomb-Scargle method} \citep{Lomb1976, Scargle1982} via $Gatspy$ software \citep{gatspy2015}. We found the rotational period of Baboquivari as $129.93 \pm 2.31$ hours \rcolor{and the amplitude of the light curve as $0.42$}. \rcolor{As seen in Fig \ref{fig:lombscargle}, the bi-modal solution was preferred as it provides a more accurate representation of the double-peaked light curve, enabling a better understanding of the asteroid's physical characteristics because the shapes of the asteroids can be approximated as a triaxial ellipsoid that results in a double-peaked light curve.} The uncertainty in the measurement was estimated using half-width at half-maximum of the peak power \citep{vanderplas2018}. After having a reliable period using only data at \rcolor{similar} phase angles, we can now apply that period to all photometric data to construct rotational phasing (see Fig. \ref{fig:lightcurve}). Because the data at different rotation phases will affect the photometric phase curve, we need to use the data at the same rotation phase or the average. As stated in \cite{shevchenko2002} that the maximum of the light curve is less affected than the minimum, we used the maximum of the light curve at around 0.25 phase of rotation (see Fig. \ref{fig:lightcurve}) to construct the photometric phase curve. Using this information, we found the difference between all absolute data and the maximum magnitude of the fit curve and added them to their reduced magnitudes. Then we found a new $H$ and $G$, constructed a new light curve and found a new maximum value. We repeated this process until $H$ and $G$ converged to a constant value. In Fig. \ref{fig:HG_phasecurveanditeration}, it can be seen $H$ and $G$ values converged to $16.05\ \pm 0.05$ and $0.22\ \pm 0.02$, respectively, as well as the photometric phase curve.

\subsection{Spectroscopy}\label{subsec:spectroscopy}
Spectral observations of the asteroid were carried out on the RTT150 telescope using the TFOSC instrument on 23 September 2019 at the phase angle 30.8\textdegree. The quantum efficiency of the CCD detector is 90\% or higher in the wavelength range from 4000\,\AA~to 8500\,\AA. Grism 15 was used as a disperser together with a slit width of 0.200 mm (3.6 arcsecs). Three low-resolution spectra with an exposure of 600 seconds each were obtained in the wavelength range from 4000\,\AA ~to 7800\,\AA ~ with a spectral resolution of 25\,\AA\ \rcolor{(see \ref{sec:appendixB}). Atmospheric conditions were spectrophotometric. In order to keep the image of the asteroid in the same slit area while obtaining its spectrum, a non-sidereal tracking option was used depending on the target's sky motion. A combined spectrum was obtained from the three spectra utilizing a median filter in case another source enter the slit and contaminate the spectra.} \rcolor{At the same night, the solar twin HIP 100963 \citep{takeda2009} was also observed in the moments of its upper culmination with identical spectral resolution and range. The reflectance spectrum of the Babaquivari was constructed by division to the solar twin spectrum with the flux normalisation at the 5500\AA\ spectral region. Because the airmass values of the asteroid and solar twin differ no more than 0.15 and the atmospheric absorption coefficient of the observatory sky in $V$-band is about 0.15 mag/airmass, which can produce a discrepancy in reflectance spectrum less than 2\%,  the atmospheric correction was not performed.} 
\begin{figure}
\centering
\includegraphics[keepaspectratio=true,scale=0.7]{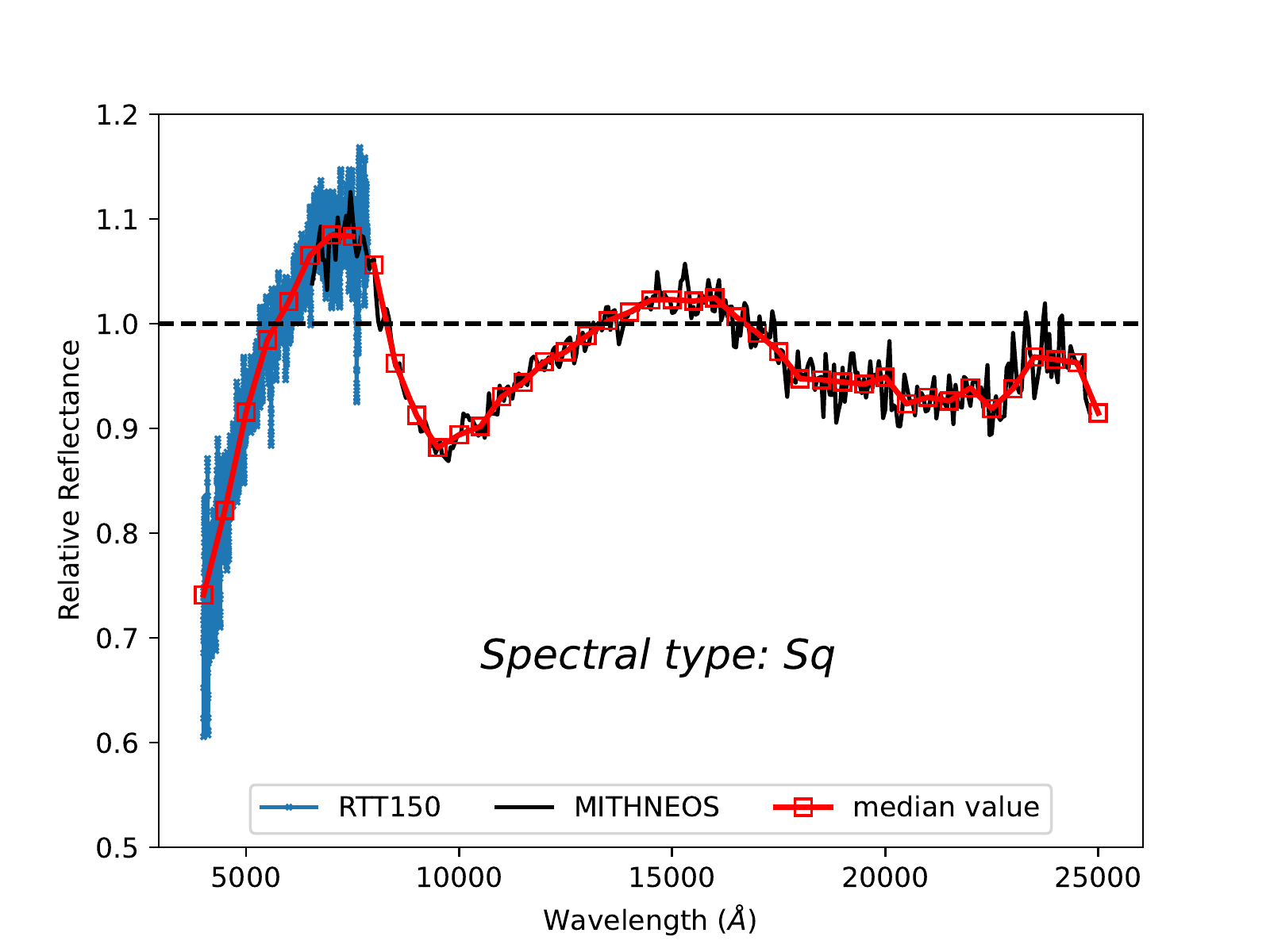}
\caption{The combined reflectance spectrum of Baboquivari, obtained from observations made on RTT150 on September 23, 2019, and MITHNEOS on 26 September 2019, is shown. MITHNEOS data normalized to that of $\sim$7000\,\AA\ where it overlaps with RTT150 spectra.  The phase angle of the object at the moment of observations corresponded to 30.8 degrees and 32 degrees. The best fit (red line) determines the spectral class of the asteroid as Sq. The red squares indicate the median values of the spectrum in the intervals used in the fitting (see the text).}
\label{fig:spec}
\end{figure}
Using RTT150 VIS spectra, the spectral class of the Baboquivari was determined based on the best fit of the reflectance spectrum by the least squares method using the median values from the 7 intervals (4400\,\AA, 5000\,\AA, 5500\,\AA, 6000\,\AA, 6500\,\AA, 7000\,\AA\ and 7500\,\AA), which were calculated within $\pm250\,$\AA\ around the central value, of the VIS spectrum taken from Table III by \cite{bus2002}. As a result, the spectral class of the Baboquivari was determined as an Sq-type in the Bus-DeMeo taxonomy. We also combined MITHNEOS IR and RTT150 VIS spectra. This time we used the same method with 42 intervals to find the best fit and obtained the same result. In Fig. \ref{fig:spec}, the combined  reflectance spectrum of the Baboquivari with the best fit (red line) is shown. The red squares indicate the median values of the spectrum in the intervals used in the fitting.

\subsection{Polarimetry}\label{subsec:polarimetry}
\begin{figure}
\centering
\includegraphics[keepaspectratio=true,scale=0.7]{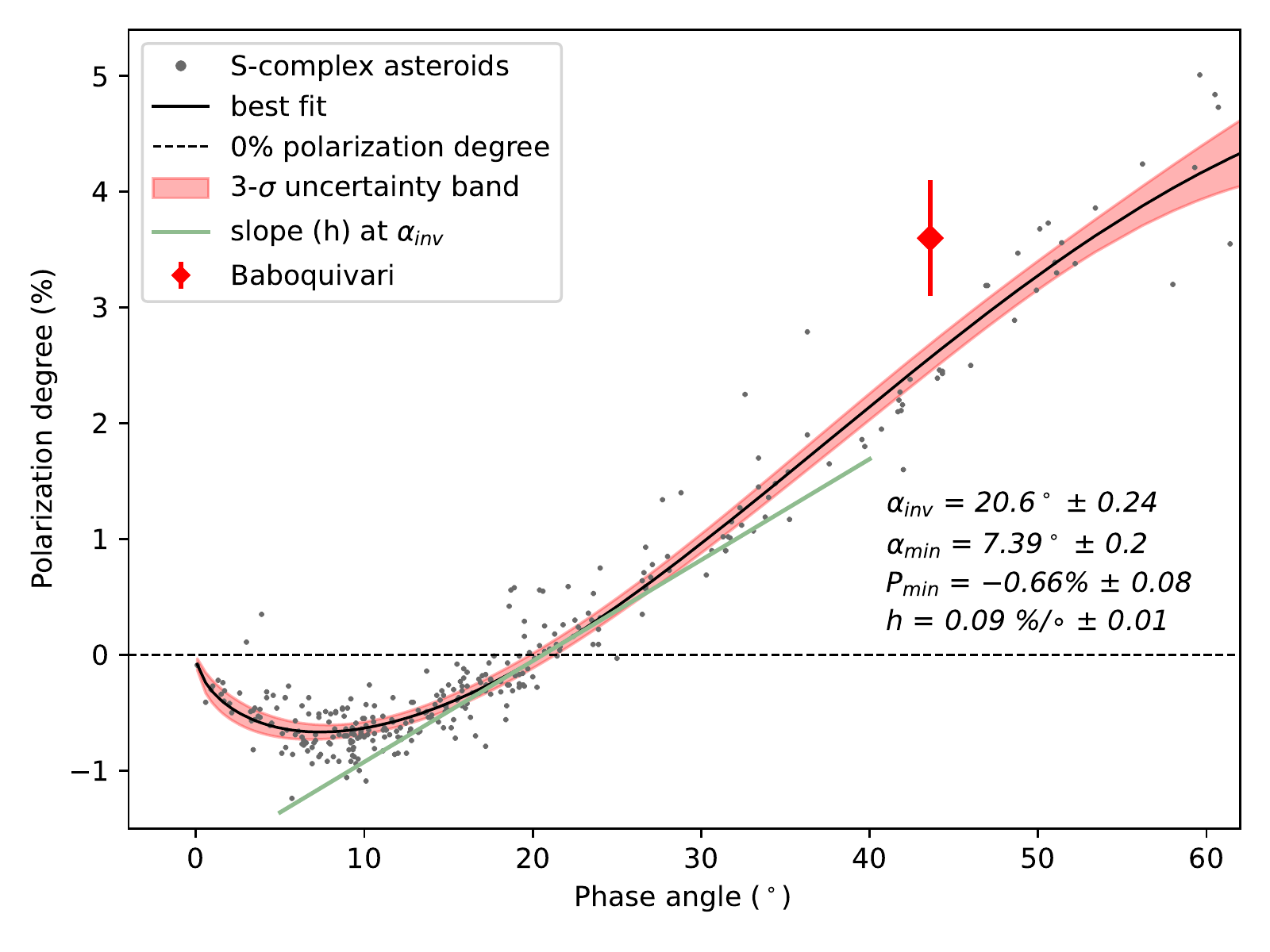}
\caption{Polarimetric phase curve for S-complex of asteroids in filter V and the location of Baboquivari. Grey dots correspond to data from the Asteroid Polarimetric Database (APD) \citep{lupishko2022}. The solid line is the best fit by the trigonometric function of \cite{lumme1993}. The green line is the slope $h$ of the curve at the inversion angle. Red diamond with error is our measurement of Baboquivari.}
\label{fig:poldegree}
\end{figure}

The geometric albedo of Baboquivari was estimated for the first time using the polarimetric method. This method is based on the dependence of the degree of polarization $P$ of the sunlight scattered off of the asteroid's surface at the phase angle $\alpha$. Phase polarimetric curve can be obtained using the trigonometric function suggested by \cite{lumme1993},

\begin{equation}
\label{eq:polalpha}
P(\alpha) = b\ sin^{c_1}(\alpha)\ cos^{c_2}(\alpha/2)\ sin(\alpha-\alpha_{inv})
\end{equation}
where $b$, $c_{1}$, $c_{2}$, and $\alpha_{inv}$ are the wavelength-dependent free parameters shaping the curve. Since we have a single point, we could not construct a polarimetric phase curve of Baboquivari. However, we compared Baboquivari with the polarimetric data of other known S-complex asteroids from the Asteroid Polarimetric Database (APD) \citep{lupishko2022} applying Eq. \eqref{eq:polalpha}, which can be seen in Fig. \ref{fig:poldegree}. \rcolor{The reason we used all S-complex asteroids is to enlarge dataset because there are only two Sq type of asteroids in APD.} We also calculated for S-complex asteroids $\alpha_{inv} = 20.6 ^{\circ} \pm\ 0.24$, $\alpha_{min} = 7.9 ^{\circ} \pm\ 0.2$, $P_{min} = -0.66\ \pm\ 0.08$ and the polarimetric slope at $\alpha_{inv}\ h = 0.09\ \%/^{\circ} \pm\ 0.01$. The geometric albedo $p_V$  can be estimated on the basis of parameters of the phase polarimetric curve, which has logarithmic dependence determined empirically \citep{bowellzellner1974, zellneretal1977, lupishko1996, cellino1999, cellino2012, cellino2014, masiero2012, shestopalov2014}. The so-called slope-albedo relation can be described as follows: 

\begin{equation}
\label{eq:polslope}
\log_{10}(p_{V}) = C_1 \log_{10}(h) + C_2
\end{equation}
where $C_1$ and $C_2$ are constants. We measured the slope of the polarimetric curve $h$ in the region of the inversion angle $\alpha_{inv}$ based on a method with a limited number of polarimetric observations at a large phase angle and the use of a low-resolution polarimeter \citep{khamitov2020}.

Our approach is as follows; the main belt asteroids can be observed from Earth on phase angles \rcolor{less} than 30\textdegree. This imposes requirements on the accuracy of polarimeters, where it is necessary to measure sufficiently small values to determine $h$, and also extremely costly in terms of observational time. Conversely, it is possible to observe NEAs at high-phase angles. Their phase angles change significantly during their close approach to Earth. Also, within 2\textdegree, the mean inversion angle $\alpha_{inv}$ for most asteroids is of the order of 19\textdegree~ \citep{mishchenko2010}. The empirical model for a polarization phase curve \cite{lumme1993} shows that the interval between the $\alpha_{inv}$ up to phase angles of 40-60\textdegree~  is close to linear. Thus, by measuring the degree of polarization at a high phase angle, it is possible to estimate the polarimetric slope as the ratio of the degree of polarization to the difference between the phase angle at which the measurement was made and the average inversion angle.

\begin{figure}[H]
\centering
\begin{subfigure}{0.48\linewidth}
    \includegraphics[width=\linewidth, height=0.9\linewidth]{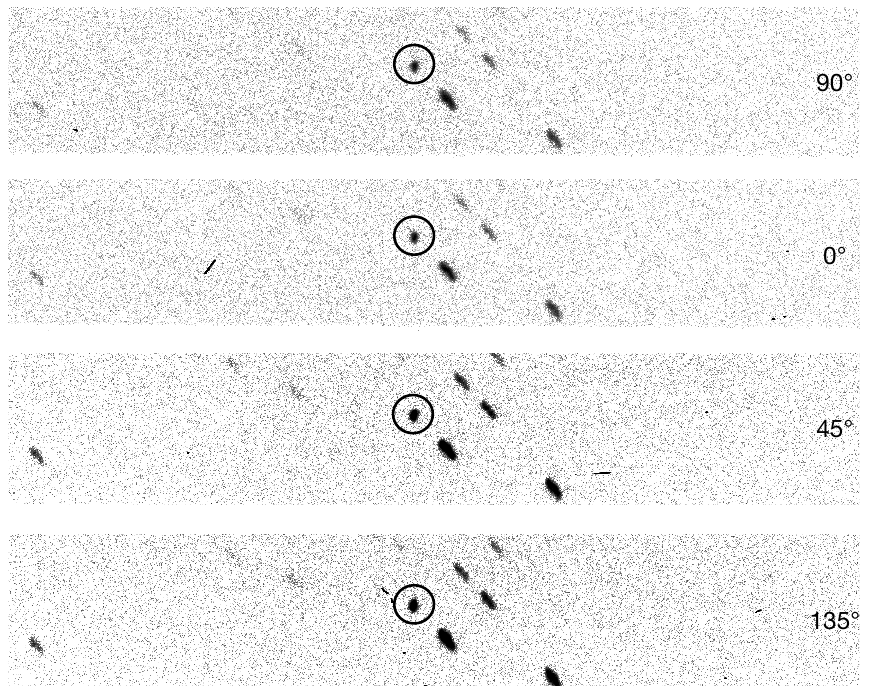}
\caption{Single exposure image}
    \label{fig:single_pol_image}
\end{subfigure}
    \hfill
\begin{subfigure}{0.48\linewidth}
    \includegraphics[width=\linewidth, height=0.9\linewidth]{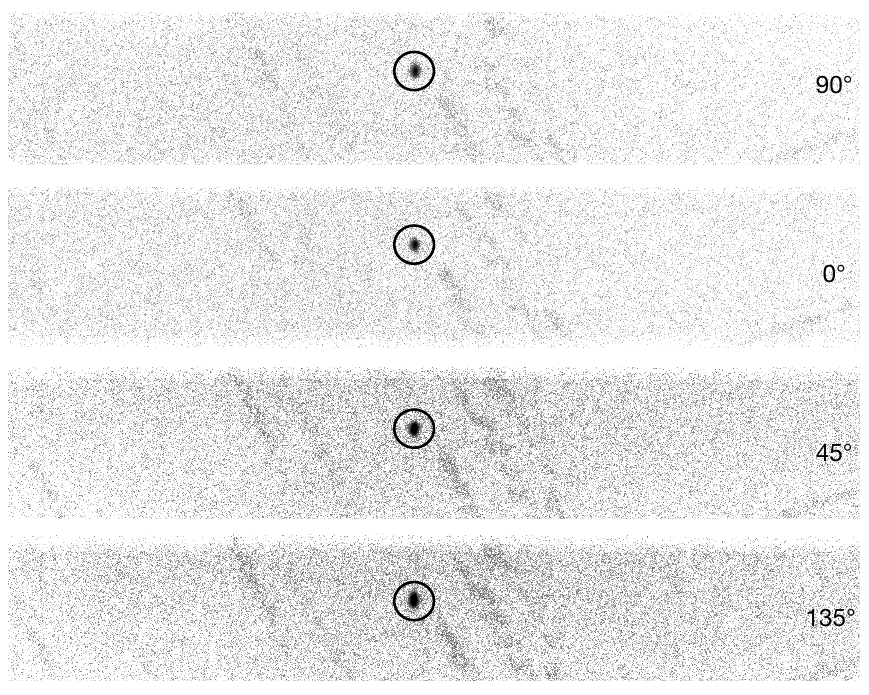}
\caption{Median combined image}
    \label{fig:median_pol_image}
\end{subfigure}
\caption{\rcolor{Polarimetric images of Baboquivari at 4 channels (90\textdegree, 0\textdegree, 45\textdegree, 135\textdegree) obtained with TFOSC-WP polarimeter. Black circles indicate Baboquivari on each channel.}}
\label{fig:tfoscwp}
\end{figure}

Our observations in the $V$-band were performed on 20 October 2019 (see Table \ref{tab:t100obs-details}) using a TFOSC-WP polarimeter with 0.2\% of the systematic error of the polarization degree. \rcolor{TFOSC-WP polarimeter allows measuring the signals of polarized light from 4 channels simultaneously: $0^\circ,45^\circ,90^\circ$ and $135^\circ$. On the base of the registered fluxes, three parameters of the Stokes vector $\vec{S}(I, Q, U)$ can be determined. Thus, the positional angle and degree of linear polarization can be estimated for a single-shot polarimetric measurement. We obtained 5 polarimetric images of Baboquivari of 300 seconds each. In Fig. \ref{fig:tfoscwp}a, an example of the single polarimetric exposure with 4 images corresponding to the different polarimetric channels is presented. 

The asteroid appears star-like, and the neighbouring stars appear elongated in the direction of the proper motion of Baboquivari due to the use of non-sidereal tracking mode. To calculate the polarimetric fluxes, a median combined image was formed by merging 5 images, which is not affected by neighbouring sources, as shown in Fig. \ref{fig:tfoscwp}b. The degree of linear polarization and the error estimation was calculated as described in \cite{helhel2015}. For calibration and analysis of instrumental polarization, the unpolarized stars (HD14069, HD21447) and strongly polarized stars (HD19820, HD25443) \citep{schmidt1992} were observed on the same night.}

We have found that the polarization degree at the phase angle 43.6\textdegree\ is $3.6\% \pm 0.5$, and we searched the polarimetric slope ($h$) of Baboquivari within 2\textdegree\ for the inversion angle of 20.6\textdegree\ and found $0.16\pm 0.03$. \rcolor{Using $C_1 = -0.8 \pm 0.041$ and $C_2 = -1.467 \pm 0.037$ values in Eq. \eqref{eq:polslope} for $p_{V} \ge 0.08$ given by \cite{cellino2015}}, we estimated the geometric albedo of the studied asteroid to be $p_{V}=0.15 \pm 0.03$.

\section{Results and Discussion}\label{sec:resanddis}
\begin{figure}
\centering
\includegraphics[keepaspectratio=true,scale=0.7]{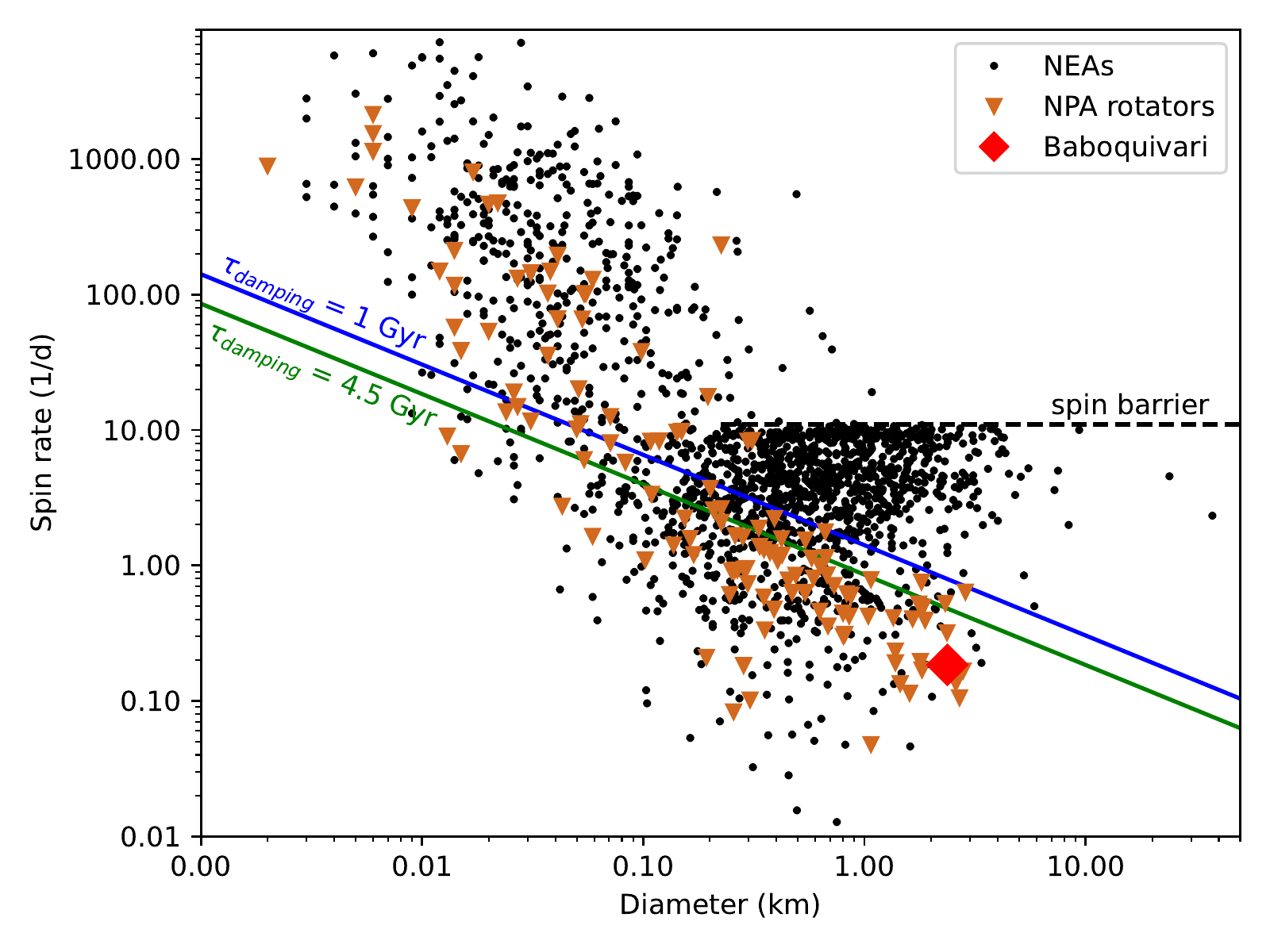}
\caption{Diameter against spin rate for near-Earth asteroids taken from Asteroid Lightcurve Database\protect\footnotemark\ \citep{warner2009}. Black dots represent NEAs (U > 2), while browns are all known or possible NPA rotators (T or T+) among NEAs. The red diamond indicates the position of Baboquivari.}
\label{fig:npa}
\end{figure}

\rcolor{The light curve and, accordingly, the rotational period of Baboquivari were constructed based on ALCDEF data together with our T100 observations; hence we determined almost the same photometric results as in \cite{warnerstephens2020}. All the periods in the literature are not at all different from each other anyway. Minor differences in periods may be due to data quality, calculation method or the use of data from different phase angles. However, all the periods clearly indicate that Baboquivari is a slow-rotating asteroid. Besides, according to a well-known relation between the diameter ($D_{eff}$, in km), absolute magnitude ($H$) and geometric albedo ($p_{V}$) as seen in Eq. \eqref{eq:dhp}, $D_{eff}$ of Baboquivari is $2.12 \pm 0.21$ km, indicating that it is relatively larger among NEAs (see Fig. \ref{fig:npa}). The variation in diameter and albedo between this work and \cite{masiero2021} could be attributed to differences in the observation methods utilized.}
 
\begin{equation}
\label{eq:dhp}
\log_{10}D_{eff} = 3.1236 - 0.2H - 0.5\log_{10}(p_{V})
\end{equation}

The lower limit of axis ratio can be calculated using Eq. \eqref{eq:abratio}, which is given by  \cite{zappala1990} where A($\alpha$) is the amplitude measured at solar phase angle $\alpha$. They also found that the average m-value is 0.03 for S-type asteroids. Using this m-value and properties derived from Fig. \ref{fig:lightcurve}b, which covers phase angle between 15\textdegree and 19.9\textdegree, we found the lower limit of the axis ratio for Baboquivari as $1.44$.

\begin{equation}
\label{eq:abratio}
a/b = 10^{0.4A(\alpha) / (1+m\alpha)}
\end{equation}

\footnotetext{LCDB, \url{https://minplanobs.org/MPInfo/php/lcdb.php}}

 \rcolor{Contrary to probable spectral types obtained using only MITHNEOS IR data, the combined data (RTT150 VIS + MITHNEOS IR) match only Sq-type in the Bus-DeMeo taxonomy which is the same as we calculated with RTT150 VIS spectra alone. Besides, the colour index V-R that we found as $0.45 \pm 0.02$ is consistent with that spectral type according to Tholen's taxonomic class colour table provided by \cite{dandy2003}. The table gives the V-R value of 0.474 for the S-spectral type of asteroids, which is comparable to our result.}

We have not seen any peculiarities in polarimetric and spectroscopic observations. However, substantial data scattering is seen in the light curve. We considered two effects that can cause this: 1) NPA rotation and 2) cometary activity. Due to the lack of observational data, the spin state and shape model of Baboquivari could not be constructed. However, significant signs of light curve suggest the asteroid could be an NPA rotator (tumbler). We considered the damping timescale ($\tau_{damping}$) of a tumbling asteroid to strengthen this situation. This is the timescale of damping the somehow excited asteroid rotation (e.g. collision, YORP) so the transition from NPA rotation to principal-axis-rotation (PA). 

\begin{equation}
\label{eq:dampingtimescale}
\tau_{damping} = \frac{P^3}{C^3 D^2}
\end{equation}
Using Eq. \eqref{eq:dampingtimescale} where $C$ is a constant value of 36, and D is the diameter in km, P is period in hours, and $\tau_{damping}$ is in Gyr \citep{pravec2014}, we calculated the damping timescale for Baboquivari is $\sim$8.5\ Gyr which is almost twice the age of the Solar system. In Fig. \ref{fig:npa}, it is evident that the damping timescales of relatively big NPA rotators ($\ge$ 0.3 km) among NEOs are more significant than $1\ Gyr$. This timescale may be returning to the beginning due to the YORP effect. The rotation of the asteroid may be slowed down by the YORP effect and may eventually turn into a tumbler \citep{vokrouhlicky2007}. The YORP timescale reaching the onset of tumbling caused by the YORP effect can be calculated using Eq. \eqref{eq:yorptimescale}.

\begin{equation}
\label{eq:yorptimescale}
\tau_{YORP} = \tau_0\left(\frac{D}{D_0}\right)^{\!2} \left(\frac{a}{a_0}\right)^{\!2}
\end{equation}
\cite{vokrouhlicky2007} gives $\tau_{0}$ as 5.3 Kyr using sampled data $D_{0} = 50\ m$, $a_{0} = 2\ AU$ and a reference initial rotation period $P_{0} = 6\ h$. It should be noted that the estimated YORP timescale would be shorter if the initial rotation period was longer than $P_{0}$. We estimated $\tau_{YORP}$ for Baboquivari as $20\ Myr$, which is quite enough to remain the asteroid in the NPA state compared to $\tau_{damping}$.

We have seen no evidence of cometary activity from \rcolor{our observations} of Baboquivari. However, more observations are needed to search for significant reflectance variations at different times. Since Baboquivari moves in a relatively eccentric orbit with $e = 0.53$, we calculated its Tisserand parameter with respect to Jupiter \citep{kresak1982, kosai1992} which is defined by,

\begin{equation}
\label{eq:tisserand}
T_{J} = \frac{a_J}{a} + 2\left[(1-e^{\!2}) \frac{a}{a_J}\right]^{\!1/2}cos(i)
\end{equation}
where $a$, $e$, and $i$ are the semimajor axis, eccentricity, and inclination of the orbit, respectively, while $a_J = 5.2\ AU$ is the semimajor axis of the orbit of Jupiter. $T_J$ of Babouqivari is 3.154. According to \cite{jewitt2012}, active asteroids have $T_J > 3.08$, so we believe Baboquivari should also be examined from this perspective in future observations.

\section*{Declaration of competing interest
}
The authors declare that they have no known competing financial interests or personal relationships that could have appeared to influence the work reported in this paper.

\section*{Acknowledgements}
We thank TÜBİTAK National Observatory for a partial support in using T100 telescope with project numbers 21AT100-1783 and 19AT100-1482. We thank TÜBİTAK National Observatory, IKI and KFU for partial support in using RTT150 (Russian - Turkish 1.5-m telescope in Antalya).
We also thank the on-duty observers and technical staff members at TÜBİTAK National Observatory for their support before and during the observations.
IMKh, IFB, RIG, and ENI thank the Academy of Sciences of the Republic of Tatarstan for their support, and this work was partially supported by the subsidy FZSM-2023-0015 allocated to the Kazan Federal University for the State assignment in the sphere of scientific activities.
This work uses data obtained from the Asteroid Lightcurve Data Exchange Format (ALCDEF) database, which is supported by funding from NASA grant 80NSSC18K0851.
Thanks to H. Aziz Kayıhan for the significant contribution and proofreading the article.
Finally, the authors would like to sincerely thank the two anonymous reviewers for their constructive and detailed feedback in the development of this manuscript.
\newpage
\rcolor{
\appendix
\section{The effect of CCD surface to magnitudes}\label{sec:appendixA}
As seen in Fig. \ref{fig:zeropoint} below, we have obtained a relation between magnitude and the distance to the centre of the image for the T100 telescope after applying all the pre-calibrations, e.g., flat field and bias correction. The linearity of this distribution is because of the characteristic optical effect on the T100 telescope CCD surface. After applying linear fit to the distribution, zero-point magnitudes can be obtained to correct instrumental magnitudes depending on the asteroid’s or an object’s distance to the CCD centre.

\begin{figure}[H]
\centering
\includegraphics[keepaspectratio=true,scale=1]{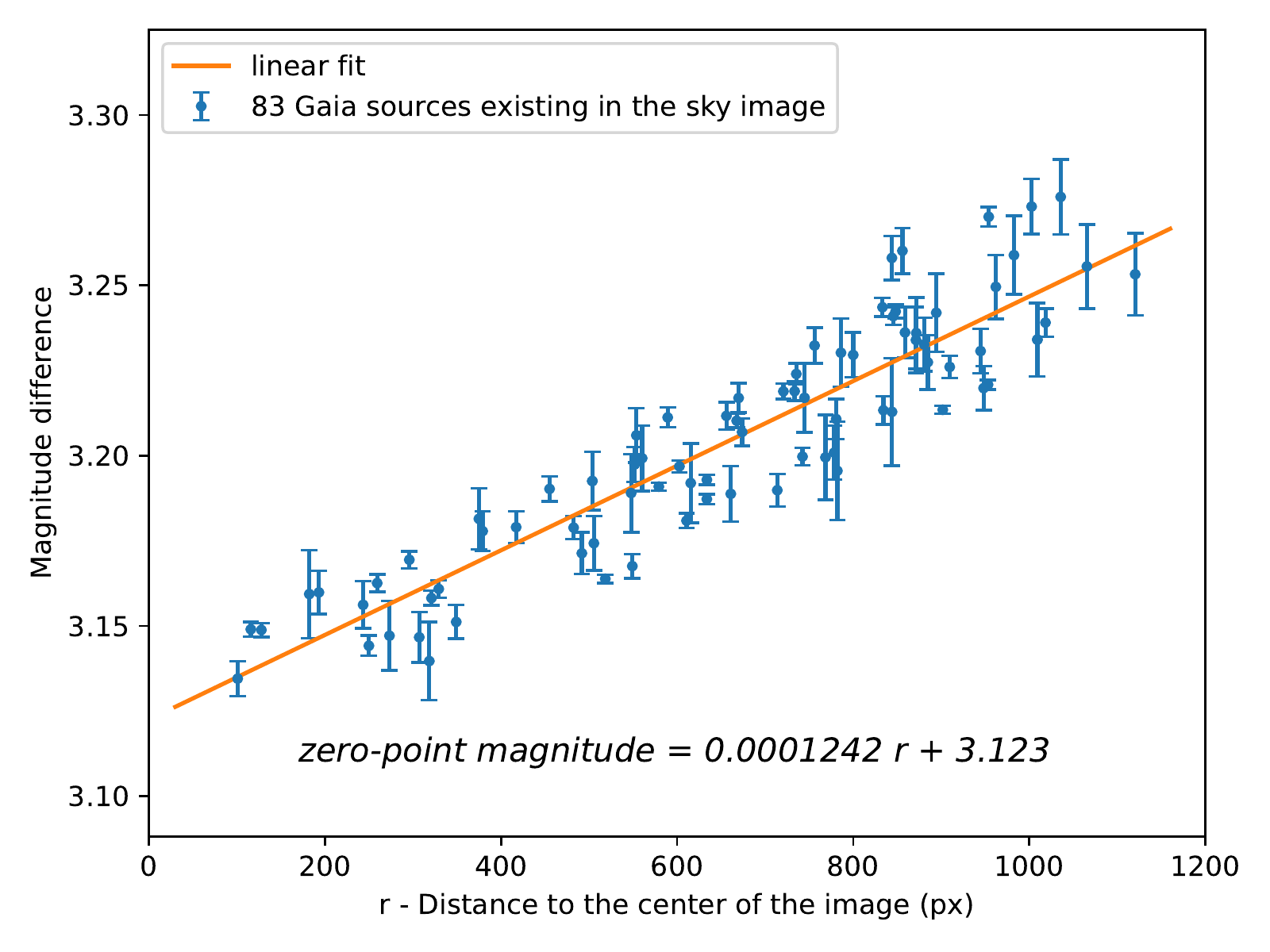}
\caption{\rcolor{Calculation of zero-point magnitude as a function of distance to the centre. The Y-axis indicates the difference between instrumental magnitude and transformed $V$-magnitude using Eq.\eqref{eq:magtransformation} while X-axis indicates the distance from the centre of the image in pixel units. The number of 83 blue dots are the stars found in the $21' \times 21'$ field of the T100 telescope also exists in the Gaia DR3 catalogue. The orange line is a linear fit to those blue dots.}}
\label{fig:zeropoint}
\end{figure}

\section{Individual spectra of (2059) Baboquivari}\label{sec:appendixB}
\begin{figure}[H]
\centering
\includegraphics[keepaspectratio=true,scale=1]{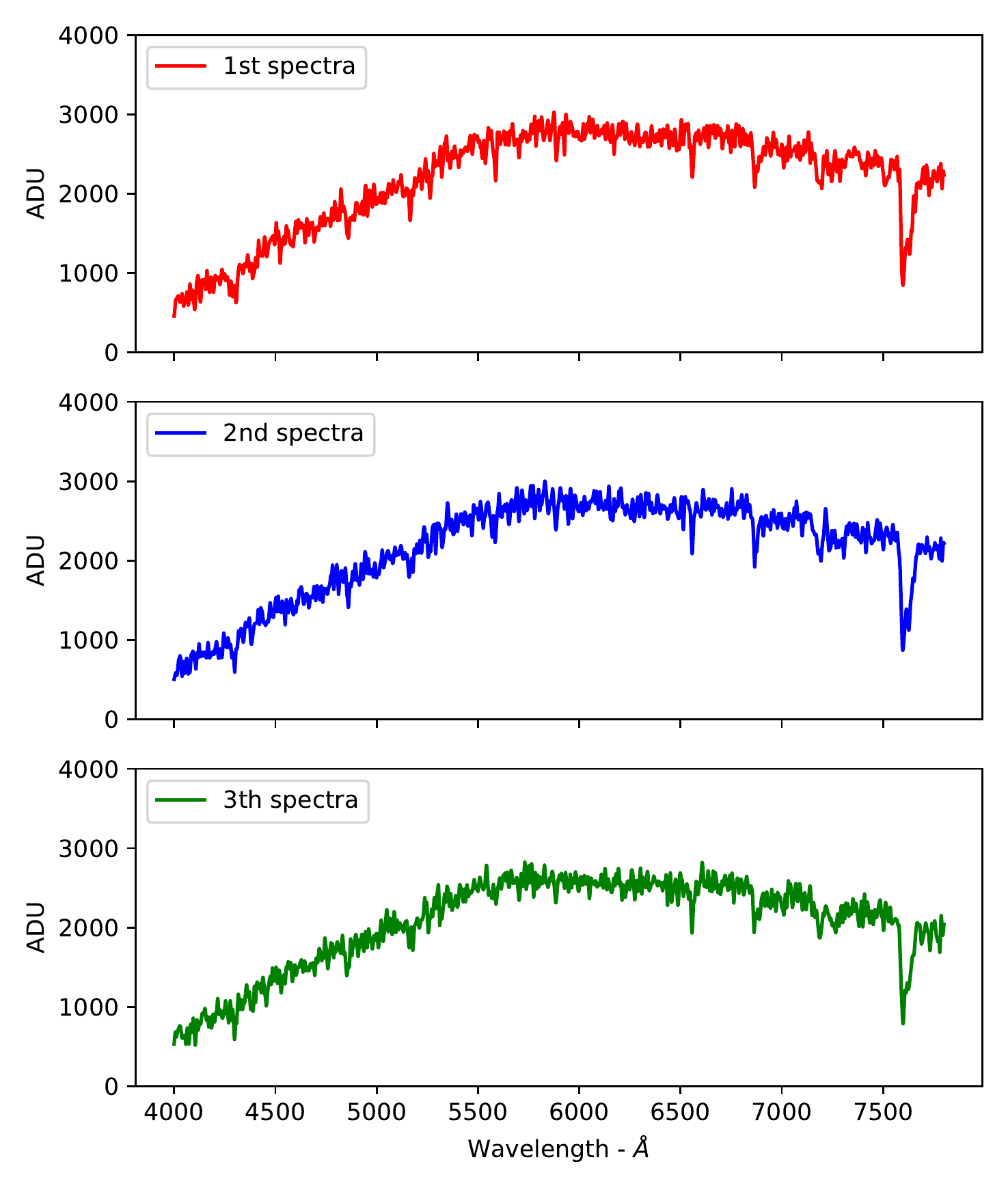}
\caption{\rcolor{Three individual spectra of Baboquivari from RTT150 telescope with 600 seconds of exposure were taken on 23 September 2019 at the phase angle 30.8\textdegree. The median combination of those three spectra was used for analysing (see Section \ref{subsec:spectroscopy}).}}
\label{fig:spectra2}
\end{figure}
}
\section*{Data availability}
Data will be made available upon request.

\bibliographystyle{model3-num-names}

\bibliography{all-references}

\begin{thebibliography}{55}
\providecommand{\natexlab}[1]{#1}
\providecommand{\url}[1]{\texttt{#1}}
\providecommand{\href}[2]{#2}
\providecommand{\path}[1]{#1}
\providecommand{\eprint}[1]{\href{http://arxiv.org/abs/#1}{\path{#1}}}
\providecommand{\DOIprefix}{doi:}
\providecommand{\ArXivprefix}{arXiv:}
\providecommand{\URLprefix}{URL: }
\providecommand{\Pubmedprefix}{pmid:}
\providecommand{\doi}[1]{\href{http://dx.doi.org/#1}{\path{#1}}}
\providecommand{\Pubmed}[1]{\href{pmid:#1}{\path{#1}}}
\providecommand{\BIBand}{and}
\providecommand{\bibinfo}[2]{#2}
\ifx\xfnm\undefined \def\xfnm[#1]{\unskip,\space#1}\fi
\bibitem[{{Aslan} et~al.(2001){Aslan}, {Bikmaev}, {Vitrichenko}, {Gumerov},
  {Dembo}, {Kamus} et~al.}]{AslanRTT150}
\bibinfo{author}{{Aslan}\xfnm[ Z.]}, \bibinfo{author}{{Bikmaev}\xfnm[ I.F.]},
  \bibinfo{author}{{Vitrichenko}\xfnm[ {\'E}.A.]},
  \bibinfo{author}{{Gumerov}\xfnm[ R.I.]}, \bibinfo{author}{{Dembo}\xfnm[
  L.A.]}, \bibinfo{author}{{Kamus}\xfnm[ S.F.]}, et~al.
\newblock \bibinfo{title}{{Preliminary Results of the Alignment and Hartmann
  Tests of the AZT-22 Telescope}}.
\newblock \bibinfo{journal}{Astronomy Letters}
  \bibinfo{year}{2001};\bibinfo{volume}{27}:\bibinfo{pages}{398--403}.
\newblock \DOIprefix\doi{10.1134/1.1374679}.
\bibitem[{{Astropy Collaboration} et~al.(2018){Astropy Collaboration},
  {Price-Whelan}, {Sip{\H{o}}cz}, {G{\"u}nther}, {Lim}, {Crawford}
  et~al.}]{astropy2018}
\bibinfo{author}{{Astropy Collaboration}\xfnm[]},
  \bibinfo{author}{{Price-Whelan}\xfnm[ A.M.]},
  \bibinfo{author}{{Sip{\H{o}}cz}\xfnm[ B.M.]},
  \bibinfo{author}{{G{\"u}nther}\xfnm[ H.M.]}, \bibinfo{author}{{Lim}\xfnm[
  P.L.]}, \bibinfo{author}{{Crawford}\xfnm[ S.M.]}, et~al.
\newblock \bibinfo{title}{{The Astropy Project: Building an Open-science
  Project and Status of the v2.0 Core Package}}.
\newblock \bibinfo{journal}{Astronomical Journal}
  \bibinfo{year}{2018};\bibinfo{volume}{156}(\bibinfo{number}{3}):\bibinfo{eid}{123}.
\newblock \DOIprefix\doi{10.3847/1538-3881/aabc4f}.
  \href{http://arxiv.org/abs/1801.02634}{\tt arXiv:1801.02634}.
\bibitem[{{Astropy Collaboration} et~al.(2013){Astropy Collaboration},
  {Robitaille}, {Tollerud}, {Greenfield}, {Droettboom}, {Bray}
  et~al.}]{astropy2013}
\bibinfo{author}{{Astropy Collaboration}\xfnm[]},
  \bibinfo{author}{{Robitaille}\xfnm[ T.P.]}, \bibinfo{author}{{Tollerud}\xfnm[
  E.J.]}, \bibinfo{author}{{Greenfield}\xfnm[ P.]},
  \bibinfo{author}{{Droettboom}\xfnm[ M.]}, \bibinfo{author}{{Bray}\xfnm[ E.]},
  et~al.
\newblock \bibinfo{title}{{Astropy: A community Python package for astronomy}}.
\newblock \bibinfo{journal}{Astronomy and Astrophysics}
  \bibinfo{year}{2013};\bibinfo{volume}{558}:\bibinfo{eid}{A33}.
\newblock \DOIprefix\doi{10.1051/0004-6361/201322068}.
  \href{http://arxiv.org/abs/1307.6212}{\tt arXiv:1307.6212}.
\bibitem[{{Bertin, E.} and {Arnouts, S.}(1996)}]{sextractor}
\bibinfo{author}{{Bertin, E.}\xfnm[]}, \bibinfo{author}{{Arnouts, S.}\xfnm[]}.
\newblock \bibinfo{title}{Sextractor: Software for source extraction}.
\newblock \bibinfo{journal}{Astron Astrophys Suppl Ser}
  \bibinfo{year}{1996};\bibinfo{volume}{117}(\bibinfo{number}{2}):\bibinfo{pages}{393--404}.
\newblock \URLprefix \url{https://doi.org/10.1051/aas:1996164}.
  \DOIprefix\doi{10.1051/aas:1996164}.
\bibitem[{{Binzel} et~al.(2019){Binzel}, {DeMeo}, {Turtelboom}, {Bus},
  {Tokunaga}, {Burbine} et~al.}]{binzelmithneos}
\bibinfo{author}{{Binzel}\xfnm[ R.P.]}, \bibinfo{author}{{DeMeo}\xfnm[ F.E.]},
  \bibinfo{author}{{Turtelboom}\xfnm[ E.V.]}, \bibinfo{author}{{Bus}\xfnm[
  S.J.]}, \bibinfo{author}{{Tokunaga}\xfnm[ A.]},
  \bibinfo{author}{{Burbine}\xfnm[ T.H.]}, et~al.
\newblock \bibinfo{title}{{Compositional distributions and evolutionary
  processes for the near-Earth object population: Results from the MIT-Hawaii
  Near-Earth Object Spectroscopic Survey (MITHNEOS)}}.
\newblock \bibinfo{journal}{Icarus}
  \bibinfo{year}{2019};\bibinfo{volume}{324}:\bibinfo{pages}{41--76}.
\newblock \DOIprefix\doi{10.1016/j.icarus.2018.12.035}.
  \href{http://arxiv.org/abs/2004.05090}{\tt arXiv:2004.05090}.
\bibitem[{Bowell et~al.(1989)Bowell, Hapke, Domingue, Lumme, Peltoniemi and
  Harris}]{bowell1989}
\bibinfo{author}{Bowell\xfnm[ E.]}, \bibinfo{author}{Hapke\xfnm[ B.]},
  \bibinfo{author}{Domingue\xfnm[ D.]}, \bibinfo{author}{Lumme\xfnm[ K.]},
  \bibinfo{author}{Peltoniemi\xfnm[ J.]}, \bibinfo{author}{Harris\xfnm[ A.]}.
\newblock \bibinfo{title}{Asteroids ii, ed}.
\newblock \bibinfo{journal}{RP Binzel, T Gehrels, \& MS Matthews}
  \bibinfo{year}{1989};\bibinfo{volume}{524}.
\bibitem[{{Bowell} and {Zellner}(1974)}]{bowellzellner1974}
\bibinfo{author}{{Bowell}\xfnm[ E.]}, \bibinfo{author}{{Zellner}\xfnm[ B.]}.
\newblock \bibinfo{title}{{Polarizations of Asteroids and Satellites}}.
\newblock In: \bibinfo{editor}{{Gehrels}\xfnm[ T.]}, editor.
  \bibinfo{booktitle}{IAU Colloq. 23: Planets, Stars, and Nebulae: Studied with
  Photopolarimetry}. \bibinfo{year}{1974}, p. \bibinfo{pages}{381}.
\bibitem[{Bradley et~al.(2020)Bradley, Sip{\H o}cz, Robitaille, Tollerud,
  Vin{\'{\i}}cius, Deil et~al.}]{bradley2020}
\bibinfo{author}{Bradley\xfnm[ L.]}, \bibinfo{author}{Sip{\H o}cz\xfnm[ B.]},
  \bibinfo{author}{Robitaille\xfnm[ T.]}, \bibinfo{author}{Tollerud\xfnm[ E.]},
  \bibinfo{author}{Vin{\'{\i}}cius\xfnm[ Z.]}, \bibinfo{author}{Deil\xfnm[
  C.]}, et~al.
\newblock \bibinfo{title}{astropy/photutils: 1.0.0}.
\newblock \bibinfo{year}{2020}.
\newblock \URLprefix \url{https://doi.org/10.5281/zenodo.4044744}.
  \DOIprefix\doi{10.5281/zenodo.4044744}.
\bibitem[{{Bus} and {Binzel}(2002)}]{bus2002}
\bibinfo{author}{{Bus}\xfnm[ S.J.]}, \bibinfo{author}{{Binzel}\xfnm[ R.P.]}.
\newblock \bibinfo{title}{{Phase II of the Small Main-Belt Asteroid
  Spectroscopic Survey. A Feature-Based Taxonomy}}.
\newblock \bibinfo{journal}{Icarus}
  \bibinfo{year}{2002};\bibinfo{volume}{158}:\bibinfo{pages}{146--177}.
\newblock \DOIprefix\doi{10.1006/icar.2002.6856}.
\bibitem[{{Carrasco} and {Bellazzini}(2022)}]{gaia_mag_transform2022}
\bibinfo{author}{{Carrasco}\xfnm[ J.M.]}, \bibinfo{author}{{Bellazzini}\xfnm[
  M.]}.
\newblock \bibinfo{title}{Relationships with other photometric systems}.
\newblock
  \bibinfo{howpublished}{\url{https://gea.esac.esa.int/archive/documentation/GDR3/Data_processing/chap_cu5pho/cu5pho_sec_photSystem/cu5pho_ssec_photRelations.html\#\$Ch5.T8e}};
  \bibinfo{year}{2022}.
\newblock \bibinfo{note}{[Online; accessed 29-January-2023]}.
\bibitem[{{Cellino} and {Bagnulo}(2014)}]{cellino2014}
\bibinfo{author}{{Cellino}\xfnm[ A.]}, \bibinfo{author}{{Bagnulo}\xfnm[ S.]}.
\newblock \bibinfo{title}{{Asteroid Polarimetry: recent advances.}}
\newblock In: \bibinfo{booktitle}{European Planetary Science Congress};
  vol.~\bibinfo{volume}{9}. \bibinfo{year}{2014}, p.
  \bibinfo{pages}{EPSC2014--171}.
\bibitem[{{Cellino} et~al.(2015){Cellino}, {Bagnulo}, {Gil-Hutton}, {Tanga},
  {Ca{\~n}ada-Assandri} and {Tedesco}}]{cellino2015}
\bibinfo{author}{{Cellino}\xfnm[ A.]}, \bibinfo{author}{{Bagnulo}\xfnm[ S.]},
  \bibinfo{author}{{Gil-Hutton}\xfnm[ R.]}, \bibinfo{author}{{Tanga}\xfnm[
  P.]}, \bibinfo{author}{{Ca{\~n}ada-Assandri}\xfnm[ M.]},
  \bibinfo{author}{{Tedesco}\xfnm[ E.F.]}.
\newblock \bibinfo{title}{{On the calibration of the relation between geometric
  albedo and polarimetric properties for the asteroids}}.
\newblock \bibinfo{journal}{Monthly Notices of the Royal Astronomical Society}
  \bibinfo{year}{2015};\bibinfo{volume}{451}(\bibinfo{number}{4}):\bibinfo{pages}{3473--3488}.
\newblock \DOIprefix\doi{10.1093/mnras/stv1188}.
  \href{http://arxiv.org/abs/1506.00554}{\tt arXiv:1506.00554}.
\bibitem[{Cellino et~al.(2009)Cellino, Dell’Oro and Tedesco}]{cellino2009}
\bibinfo{author}{Cellino\xfnm[ A.]}, \bibinfo{author}{Dell’Oro\xfnm[ A.]},
  \bibinfo{author}{Tedesco\xfnm[ E.]}.
\newblock \bibinfo{title}{Asteroid families: Current situation}.
\newblock \bibinfo{journal}{Planetary and Space Science}
  \bibinfo{year}{2009};\bibinfo{volume}{57}(\bibinfo{number}{2}):\bibinfo{pages}{173--182}.
\newblock \URLprefix
  \url{https://www.sciencedirect.com/science/article/pii/S0032063308002377}.
  \DOIprefix\doi{https://doi.org/10.1016/j.pss.2008.07.028};
  \bibinfo{note}{catastrophic Disruption in the Solar System}.
\bibitem[{{Cellino} et~al.(2012){Cellino}, {Gil-Hutton}, {Dell'Oro},
  {Bendjoya}, {Ca{\~n}ada-Assandri} and {Di Martino}}]{cellino2012}
\bibinfo{author}{{Cellino}\xfnm[ A.]}, \bibinfo{author}{{Gil-Hutton}\xfnm[
  R.]}, \bibinfo{author}{{Dell'Oro}\xfnm[ A.]},
  \bibinfo{author}{{Bendjoya}\xfnm[ P.]},
  \bibinfo{author}{{Ca{\~n}ada-Assandri}\xfnm[ M.]}, \bibinfo{author}{{Di
  Martino}\xfnm[ M.]}.
\newblock \bibinfo{title}{{A new calibration of the albedo-polarization
  relation for the asteroids}}.
\newblock \bibinfo{journal}{Journal of Quantitative Spectroscopy and Radiative
  Transfer}
  \bibinfo{year}{2012};\bibinfo{volume}{113}(\bibinfo{number}{18}):\bibinfo{pages}{2552--2560}.
\newblock \DOIprefix\doi{10.1016/j.jqsrt.2012.03.010}.
\bibitem[{{Cellino} et~al.(1999){Cellino}, {Hutton}, {Tedesco}, {Di Martino}
  and {Brunini}}]{cellino1999}
\bibinfo{author}{{Cellino}\xfnm[ A.]}, \bibinfo{author}{{Hutton}\xfnm[ R.G.]},
  \bibinfo{author}{{Tedesco}\xfnm[ E.F.]}, \bibinfo{author}{{Di Martino}\xfnm[
  M.]}, \bibinfo{author}{{Brunini}\xfnm[ A.]}.
\newblock \bibinfo{title}{{Polarimetric Observations of Small Asteroids:
  Preliminary Results}}.
\newblock \bibinfo{journal}{Icarus}
  \bibinfo{year}{1999};\bibinfo{volume}{138}(\bibinfo{number}{1}):\bibinfo{pages}{129--140}.
\newblock \DOIprefix\doi{10.1006/icar.1998.6062}.
\bibitem[{{Dandy} et~al.(2003){Dandy}, {Fitzsimmons} and
  {Collander-Brown}}]{dandy2003}
\bibinfo{author}{{Dandy}\xfnm[ C.L.]}, \bibinfo{author}{{Fitzsimmons}\xfnm[
  A.]}, \bibinfo{author}{{Collander-Brown}\xfnm[ S.J.]}.
\newblock \bibinfo{title}{{Optical colors of 56 near-Earth objects: trends with
  size and orbit}}.
\newblock \bibinfo{journal}{Icarus}
  \bibinfo{year}{2003};\bibinfo{volume}{163}(\bibinfo{number}{2}):\bibinfo{pages}{363--373}.
\newblock \DOIprefix\doi{10.1016/S0019-1035(03)00087-3}.
\bibitem[{{DeMeo} et~al.(2009){DeMeo}, {Binzel}, {Slivan} and
  {Bus}}]{demeo2009}
\bibinfo{author}{{DeMeo}\xfnm[ F.E.]}, \bibinfo{author}{{Binzel}\xfnm[ R.P.]},
  \bibinfo{author}{{Slivan}\xfnm[ S.M.]}, \bibinfo{author}{{Bus}\xfnm[ S.J.]}.
\newblock \bibinfo{title}{{An extension of the Bus asteroid taxonomy into the
  near-infrared}}.
\newblock \bibinfo{journal}{Icarus}
  \bibinfo{year}{2009};\bibinfo{volume}{202}(\bibinfo{number}{1}):\bibinfo{pages}{160--180}.
\newblock \DOIprefix\doi{10.1016/j.icarus.2009.02.005}.
\bibitem[{Fischler and Bolles(1981)}]{fischler1981}
\bibinfo{author}{Fischler\xfnm[ M.A.]}, \bibinfo{author}{Bolles\xfnm[ R.C.]}.
\newblock \bibinfo{title}{Random sample consensus: a paradigm for model fitting
  with applications to image analysis and automated cartography}.
\newblock \bibinfo{journal}{Communications of the ACM}
  \bibinfo{year}{1981};\bibinfo{volume}{24}(\bibinfo{number}{6}):\bibinfo{pages}{381--395}.
\bibitem[{{Fowler} and {Chillemi}(1992)}]{FowlerChillemi}
\bibinfo{author}{{Fowler}\xfnm[ J.W.]}, \bibinfo{author}{{Chillemi}\xfnm[
  J.R.]}.
\newblock \bibinfo{title}{IRAS asteroid data processing};
  chap.~\bibinfo{chapter}{-}.
\newblock \bibinfo{address}{Hanscom AF Base, MA}: \bibinfo{publisher}{Phillips
  Laboratory}; \bibinfo{year}{1992}, p. \bibinfo{pages}{17--43}.
\bibitem[{{Gaia Collaboration} et~al.(2021){Gaia Collaboration}, {Brown},
  {Vallenari}, {Prusti}, {de Bruijne}, {Babusiaux} et~al.}]{gaia2021}
\bibinfo{author}{{Gaia Collaboration}\xfnm[]}, \bibinfo{author}{{Brown}\xfnm[
  A.G.A.]}, \bibinfo{author}{{Vallenari}\xfnm[ A.]},
  \bibinfo{author}{{Prusti}\xfnm[ T.]}, \bibinfo{author}{{de Bruijne}\xfnm[
  J.H.J.]}, \bibinfo{author}{{Babusiaux}\xfnm[ C.]}, et~al.
\newblock \bibinfo{title}{{Gaia Early Data Release 3. Summary of the contents
  and survey properties}}.
\newblock \bibinfo{journal}{Astronomy and Astrophysics}
  \bibinfo{year}{2021};\bibinfo{volume}{649}:\bibinfo{eid}{A1}.
\newblock \DOIprefix\doi{10.1051/0004-6361/202039657}.
  \href{http://arxiv.org/abs/2012.01533}{\tt arXiv:2012.01533}.
\bibitem[{Ginsburg et~al.(2019)Ginsburg, Sip{\H{o}}cz, Brasseur, Cowperthwaite,
  Craig, Deil et~al.}]{ginsburg2019}
\bibinfo{author}{Ginsburg\xfnm[ A.]}, \bibinfo{author}{Sip{\H{o}}cz\xfnm[
  B.M.]}, \bibinfo{author}{Brasseur\xfnm[ C.E.]},
  \bibinfo{author}{Cowperthwaite\xfnm[ P.S.]}, \bibinfo{author}{Craig\xfnm[
  M.W.]}, \bibinfo{author}{Deil\xfnm[ C.]}, et~al.
\newblock \bibinfo{title}{astroquery: An astronomical web-querying package in
  python}.
\newblock \bibinfo{journal}{The Astronomical Journal}
  \bibinfo{year}{2019};\bibinfo{volume}{157}(\bibinfo{number}{3}):\bibinfo{pages}{98}.
\newblock \URLprefix \url{https://doi.org/10.3847/1538-3881/aafc33}.
  \DOIprefix\doi{10.3847/1538-3881/aafc33}.
\bibitem[{{Helhel} et~al.(2015){Helhel}, {Khamitov}, {Kahya}, {Bayar}, {Kaynar}
  and {Gumerov}}]{helhel2015}
\bibinfo{author}{{Helhel}\xfnm[ S.]}, \bibinfo{author}{{Khamitov}\xfnm[ I.]},
  \bibinfo{author}{{Kahya}\xfnm[ G.]}, \bibinfo{author}{{Bayar}\xfnm[ C.]},
  \bibinfo{author}{{Kaynar}\xfnm[ S.]}, \bibinfo{author}{{Gumerov}\xfnm[ R.]}.
\newblock \bibinfo{title}{{Double-wedged Wollaston-type polarimeter design and
  integration to RTT150-TFOSC; initial tests, calibration, and
  characteristics}}.
\newblock \bibinfo{journal}{Experimental Astronomy}
  \bibinfo{year}{2015};\bibinfo{volume}{39}(\bibinfo{number}{3}):\bibinfo{pages}{595--604}.
\newblock \DOIprefix\doi{10.1007/s10686-015-9468-8}.
\bibitem[{{Jewitt}(2012)}]{jewitt2012}
\bibinfo{author}{{Jewitt}\xfnm[ D.]}.
\newblock \bibinfo{title}{{The Active Asteroids}}.
\newblock \bibinfo{journal}{Astronomical Journal}
  \bibinfo{year}{2012};\bibinfo{volume}{143}(\bibinfo{number}{3}):\bibinfo{eid}{66}.
\newblock \DOIprefix\doi{10.1088/0004-6256/143/3/66}.
  \href{http://arxiv.org/abs/1112.5220}{\tt arXiv:1112.5220}.
\bibitem[{{Khamitov} et~al.(2020){Khamitov}, {Gumerov}, {Bikmaev}, {Helhel},
  {Irtuganov}, {Melnikov} et~al.}]{khamitov2020}
\bibinfo{author}{{Khamitov}\xfnm[ I.M.]}, \bibinfo{author}{{Gumerov}\xfnm[
  R.I.]}, \bibinfo{author}{{Bikmaev}\xfnm[ I.F.]},
  \bibinfo{author}{{Helhel}\xfnm[ S.]}, \bibinfo{author}{{Irtuganov}\xfnm[
  E.N.]}, \bibinfo{author}{{Melnikov}\xfnm[ S.S.]}, et~al.
\newblock \bibinfo{title}{{Studies of physical parameters of kilometer sized
  NEA by the RTT-150 telescope}}.
\newblock In: \bibinfo{booktitle}{IAU General Assembly}. \bibinfo{year}{2020},
  p. \bibinfo{pages}{30--30}.
\newblock \DOIprefix\doi{10.1017/S174392131900334X}.
\bibitem[{{Kosai}(1992)}]{kosai1992}
\bibinfo{author}{{Kosai}\xfnm[ H.]}.
\newblock \bibinfo{title}{{Short-Period Comets and Apollo-Amor-Aten Type
  Asteroids in View of Tisserand Invariant}}.
\newblock \bibinfo{journal}{Celestial Mechanics and Dynamical Astronomy}
  \bibinfo{year}{1992};\bibinfo{volume}{54}(\bibinfo{number}{1-3}):\bibinfo{pages}{237--240}.
\newblock \DOIprefix\doi{10.1007/BF00049556}.
\bibitem[{{Kresak}(1982)}]{kresak1982}
\bibinfo{author}{{Kresak}\xfnm[ L.]}.
\newblock \bibinfo{title}{{On the Similarity of Orbits of Associated Comets,
  Asteroids and Meteoroids}}.
\newblock \bibinfo{journal}{Bulletin of the Astronomical Institutes of
  Czechoslovakia}
  \bibinfo{year}{1982};\bibinfo{volume}{33}:\bibinfo{pages}{104}.
\bibitem[{Lang et~al.(2010)Lang, Hogg, Mierle, Blanton and Roweis}]{lang2010}
\bibinfo{author}{Lang\xfnm[ D.]}, \bibinfo{author}{Hogg\xfnm[ D.W.]},
  \bibinfo{author}{Mierle\xfnm[ K.]}, \bibinfo{author}{Blanton\xfnm[ M.]},
  \bibinfo{author}{Roweis\xfnm[ S.]}.
\newblock \bibinfo{title}{{ASTROMETRY}.{NET}: {BLIND} {ASTROMETRIC}
  {CALIBRATION} {OF} {ARBITRARY} {ASTRONOMICAL} {IMAGES}}.
\newblock \bibinfo{journal}{The Astronomical Journal}
  \bibinfo{year}{2010};\bibinfo{volume}{139}(\bibinfo{number}{5}):\bibinfo{pages}{1782--1800}.
\newblock \URLprefix \url{https://doi.org/10.1088/0004-6256/139/5/1782}.
  \DOIprefix\doi{10.1088/0004-6256/139/5/1782}.
\bibitem[{{Lomb}(1976)}]{Lomb1976}
\bibinfo{author}{{Lomb}\xfnm[ N.R.]}.
\newblock \bibinfo{title}{{Least-Squares Frequency Analysis of Unequally Spaced
  Data}}.
\newblock \bibinfo{journal}{Astrophysics and Space Science}
  \bibinfo{year}{1976};\bibinfo{volume}{39}(\bibinfo{number}{2}):\bibinfo{pages}{447--462}.
\newblock \DOIprefix\doi{10.1007/BF00648343}.
\bibitem[{{Lumme} and {Muinonen}(1993)}]{lumme1993}
\bibinfo{author}{{Lumme}\xfnm[ K.]}, \bibinfo{author}{{Muinonen}\xfnm[ K.O.]}.
\newblock \bibinfo{title}{{A Two-Parameter System for Linear Polarization of
  Some Solar System Objects}}.
\newblock In: \bibinfo{booktitle}{Asteroids, Comets, Meteors 1993}; vol.
  \bibinfo{volume}{810} of \emph{\bibinfo{series}{LPI Contributions}}.
  \bibinfo{year}{1993}, p. \bibinfo{pages}{194}.
\bibitem[{{Lupishko}(2022)}]{lupishko2022}
\bibinfo{author}{{Lupishko}\xfnm[ D.]}.
\newblock \bibinfo{title}{{Asteroid Polametric Database (APD) Bundle V2.0}}.
\newblock \bibinfo{journal}{NASA Planetary Data System}
  \bibinfo{year}{2022};:\bibinfo{pages}{1}\DOIprefix\doi{10.26033/hyf9-4762}.
\bibitem[{{Lupishko} and {Mohamed}(1996)}]{lupishko1996}
\bibinfo{author}{{Lupishko}\xfnm[ D.F.]}, \bibinfo{author}{{Mohamed}\xfnm[
  R.A.]}.
\newblock \bibinfo{title}{{A New Calibration of the Polarimetric Albedo Scale
  of Asteroids}}.
\newblock \bibinfo{journal}{Icarus}
  \bibinfo{year}{1996};\bibinfo{volume}{119}(\bibinfo{number}{1}):\bibinfo{pages}{209--213}.
\newblock \DOIprefix\doi{10.1006/icar.1996.0013}.
\bibitem[{{Masiero} et~al.(2021){Masiero}, {Mainzer}, {Bauer}, {Cutri}, {Grav},
  {Kramer} et~al.}]{masiero2021}
\bibinfo{author}{{Masiero}\xfnm[ J.R.]}, \bibinfo{author}{{Mainzer}\xfnm[
  A.K.]}, \bibinfo{author}{{Bauer}\xfnm[ J.M.]}, \bibinfo{author}{{Cutri}\xfnm[
  R.M.]}, \bibinfo{author}{{Grav}\xfnm[ T.]}, \bibinfo{author}{{Kramer}\xfnm[
  E.]}, et~al.
\newblock \bibinfo{title}{{Asteroid Diameters and Albedos from NEOWISE
  Reactivation Mission Years Six and Seven}}.
\newblock \bibinfo{journal}{The Planetary Science Journal}
  \bibinfo{year}{2021};\bibinfo{volume}{2}(\bibinfo{number}{4}):\bibinfo{eid}{162}.
\newblock \DOIprefix\doi{10.3847/PSJ/ac15fb}.
  \href{http://arxiv.org/abs/2107.07481}{\tt arXiv:2107.07481}.
\bibitem[{{Masiero} et~al.(2012){Masiero}, {Mainzer}, {Grav}, {Bauer},
  {Wright}, {McMillan} et~al.}]{masiero2012}
\bibinfo{author}{{Masiero}\xfnm[ J.R.]}, \bibinfo{author}{{Mainzer}\xfnm[
  A.K.]}, \bibinfo{author}{{Grav}\xfnm[ T.]}, \bibinfo{author}{{Bauer}\xfnm[
  J.M.]}, \bibinfo{author}{{Wright}\xfnm[ E.L.]},
  \bibinfo{author}{{McMillan}\xfnm[ R.S.]}, et~al.
\newblock \bibinfo{title}{{A Revised Asteroid Polarization-Albedo Relationship
  Using WISE/NEOWISE Data}}.
\newblock \bibinfo{journal}{The Astrophysical Journal}
  \bibinfo{year}{2012};\bibinfo{volume}{749}(\bibinfo{number}{2}):\bibinfo{eid}{104}.
\newblock \DOIprefix\doi{10.1088/0004-637X/749/2/104}.
  \href{http://arxiv.org/abs/1202.2379}{\tt arXiv:1202.2379}.
\bibitem[{{Mishchenko} et~al.(2010){Mishchenko}, {Rosenbush}, {Kiselev},
  {Lupishko}, {Tishkovets}, {Kaydash} et~al.}]{mishchenko2010}
\bibinfo{author}{{Mishchenko}\xfnm[ M.I.]}, \bibinfo{author}{{Rosenbush}\xfnm[
  V.K.]}, \bibinfo{author}{{Kiselev}\xfnm[ N.N.]},
  \bibinfo{author}{{Lupishko}\xfnm[ D.F.]}, \bibinfo{author}{{Tishkovets}\xfnm[
  V.P.]}, \bibinfo{author}{{Kaydash}\xfnm[ V.G.]}, et~al.
\newblock \bibinfo{title}{{Polarimetric Remote Sensing of Solar System
  Objects}}.
\newblock \bibinfo{journal}{arXiv e-prints}
  \bibinfo{year}{2010};:\bibinfo{eid}{arXiv:1010.1171}\DOIprefix\doi{10.48550/arXiv.1010.1171}.
  \href{http://arxiv.org/abs/1010.1171}{\tt arXiv:1010.1171}.
\bibitem[{Muinonen et~al.(2010)Muinonen, Belskaya, Cellino, Delbò,
  Levasseur-Regourd, Penttilä et~al.}]{muinonen2010}
\bibinfo{author}{Muinonen\xfnm[ K.]}, \bibinfo{author}{Belskaya\xfnm[ I.N.]},
  \bibinfo{author}{Cellino\xfnm[ A.]}, \bibinfo{author}{Delbò\xfnm[ M.]},
  \bibinfo{author}{Levasseur-Regourd\xfnm[ A.C.]},
  \bibinfo{author}{Penttilä\xfnm[ A.]}, et~al.
\newblock \bibinfo{title}{A three-parameter magnitude phase function for
  asteroids}.
\newblock \bibinfo{journal}{Icarus}
  \bibinfo{year}{2010};\bibinfo{volume}{209}(\bibinfo{number}{2}):\bibinfo{pages}{542--555}.
\newblock \URLprefix
  \url{https://www.sciencedirect.com/science/article/pii/S001910351000151X}.
  \DOIprefix\doi{https://doi.org/10.1016/j.icarus.2010.04.003}.
\bibitem[{{Ortiz} et~al.(2020){Ortiz}, {Sicardy}, {Camargo}, {Santos-Sanz} and
  {Braga-Ribas}}]{ortiz2020}
\bibinfo{author}{{Ortiz}\xfnm[ J.L.]}, \bibinfo{author}{{Sicardy}\xfnm[ B.]},
  \bibinfo{author}{{Camargo}\xfnm[ J.I.B.]},
  \bibinfo{author}{{Santos-Sanz}\xfnm[ P.]},
  \bibinfo{author}{{Braga-Ribas}\xfnm[ F.]}.
\newblock \bibinfo{title}{{Stellar occultation by TNOs: from predictions to
  observations}}.
\newblock In: \bibinfo{editor}{{Prialnik}\xfnm[ D.]},
  \bibinfo{editor}{{Barucci}\xfnm[ M.A.]}, \bibinfo{editor}{{Young}\xfnm[ L.]},
  editors. \bibinfo{booktitle}{The Trans-Neptunian Solar System}.
  \bibinfo{year}{2020}, p. \bibinfo{pages}{413--437}.
\newblock \DOIprefix\doi{10.1016/B978-0-12-816490-7.00019-9}.
\bibitem[{{Pravec} et~al.(2014){Pravec}, {Scheirich}, {{\v{D}}urech},
  {Pollock}, {Ku{\v{s}}nir{\'a}k}, {Hornoch} et~al.}]{pravec2014}
\bibinfo{author}{{Pravec}\xfnm[ P.]}, \bibinfo{author}{{Scheirich}\xfnm[ P.]},
  \bibinfo{author}{{{\v{D}}urech}\xfnm[ J.]}, \bibinfo{author}{{Pollock}\xfnm[
  J.]}, \bibinfo{author}{{Ku{\v{s}}nir{\'a}k}\xfnm[ P.]},
  \bibinfo{author}{{Hornoch}\xfnm[ K.]}, et~al.
\newblock \bibinfo{title}{{The tumbling spin state of (99942) Apophis}}.
\newblock \bibinfo{journal}{Icarus}
  \bibinfo{year}{2014};\bibinfo{volume}{233}:\bibinfo{pages}{48--60}.
\newblock \DOIprefix\doi{10.1016/j.icarus.2014.01.026}.
\bibitem[{{Rayner} et~al.(2003){Rayner}, {Toomey}, {Onaka}, {Denault},
  {Stahlberger}, {Vacca} et~al.}]{rayner2003}
\bibinfo{author}{{Rayner}\xfnm[ J.T.]}, \bibinfo{author}{{Toomey}\xfnm[ D.W.]},
  \bibinfo{author}{{Onaka}\xfnm[ P.M.]}, \bibinfo{author}{{Denault}\xfnm[
  A.J.]}, \bibinfo{author}{{Stahlberger}\xfnm[ W.E.]},
  \bibinfo{author}{{Vacca}\xfnm[ W.D.]}, et~al.
\newblock \bibinfo{title}{{SpeX: A Medium-Resolution 0.8-5.5 Micron
  Spectrograph and Imager for the NASA Infrared Telescope Facility}}.
\newblock \bibinfo{journal}{Publications of the Astronomical Society of the
  Pacific}
  \bibinfo{year}{2003};\bibinfo{volume}{115}(\bibinfo{number}{805}):\bibinfo{pages}{362--382}.
\newblock \DOIprefix\doi{10.1086/367745}.
\bibitem[{{Scargle}(1982)}]{Scargle1982}
\bibinfo{author}{{Scargle}\xfnm[ J.D.]}.
\newblock \bibinfo{title}{{Studies in astronomical time series analysis. II.
  Statistical aspects of spectral analysis of unevenly spaced data.}}
\newblock \bibinfo{journal}{Astrophysical Journal}
  \bibinfo{year}{1982};\bibinfo{volume}{263}:\bibinfo{pages}{835--853}.
\newblock \DOIprefix\doi{10.1086/160554}.
\bibitem[{{Schmidt} et~al.(1992){Schmidt}, {Elston} and {Lupie}}]{schmidt1992}
\bibinfo{author}{{Schmidt}\xfnm[ G.D.]}, \bibinfo{author}{{Elston}\xfnm[ R.]},
  \bibinfo{author}{{Lupie}\xfnm[ O.L.]}.
\newblock \bibinfo{title}{{The Hubble Space Telescope Northern-Hemisphere Grid
  of Stellar Polarimetric Standards}}.
\newblock \bibinfo{journal}{The Astronomical Journal}
  \bibinfo{year}{1992};\bibinfo{volume}{104}:\bibinfo{pages}{1563}.
\newblock \DOIprefix\doi{10.1086/116341}.
\bibitem[{{Shestopalov} and {Golubeva}(2014)}]{shestopalov2014}
\bibinfo{author}{{Shestopalov}\xfnm[ D.I.]}, \bibinfo{author}{{Golubeva}\xfnm[
  L.F.]}.
\newblock \bibinfo{title}{{Polarized Light Scattered from Asteroid Surfaces. V.
  Can We Estimate Polarization Maximum for Main Belt Asteroids?}}
\newblock In: \bibinfo{booktitle}{45th Annual Lunar and Planetary Science
  Conference}. Lunar and Planetary Science Conference; \bibinfo{year}{2014}, p.
  \bibinfo{pages}{1062}.
\bibitem[{{Shevchenko} et~al.(2002){Shevchenko}, {Belskaya}, {Krugly}, {Chiomy}
  and {Gaftonyuk}}]{shevchenko2002}
\bibinfo{author}{{Shevchenko}\xfnm[ V.G.]}, \bibinfo{author}{{Belskaya}\xfnm[
  I.N.]}, \bibinfo{author}{{Krugly}\xfnm[ Y.N.]},
  \bibinfo{author}{{Chiomy}\xfnm[ V.G.]}, \bibinfo{author}{{Gaftonyuk}\xfnm[
  N.M.]}.
\newblock \bibinfo{title}{{Asteroid Observations at Low Phase Angles. II. 5
  Astraea, 75 Eurydike, 77 Frigga, 105 Artemis, 119 Althaea, 124 Alkeste, and
  201 Penelope}}.
\newblock \bibinfo{journal}{Icarus}
  \bibinfo{year}{2002};\bibinfo{volume}{155}(\bibinfo{number}{2}):\bibinfo{pages}{365--374}.
\newblock \DOIprefix\doi{10.1006/icar.2001.6651}.
\bibitem[{{Sonka} et~al.(2021){Sonka}, {Birlan}, {Nedelcu}, {Badescu} and
  {Henriksen}}]{sonka2021}
\bibinfo{author}{{Sonka}\xfnm[ A.]}, \bibinfo{author}{{Birlan}\xfnm[ M.]},
  \bibinfo{author}{{Nedelcu}\xfnm[ D.A.]}, \bibinfo{author}{{Badescu}\xfnm[
  O.]}, \bibinfo{author}{{Henriksen}\xfnm[ A.I.]}.
\newblock \bibinfo{title}{{Photometric observations of Low Delta v asteroids}}.
\newblock \bibinfo{journal}{Romanian Astronomical Journal}
  \bibinfo{year}{2021};\bibinfo{volume}{31}(\bibinfo{number}{3}):\bibinfo{pages}{293--302}.
\bibitem[{{Stetson}(1987)}]{irafstetson1987}
\bibinfo{author}{{Stetson}\xfnm[ P.B.]}.
\newblock \bibinfo{title}{{DAOPHOT: A Computer Program for Crowded-Field
  Stellar Photometry}}.
\newblock \bibinfo{journal}{Publications of the Astronomical Society of the
  Pacific} \bibinfo{year}{1987};\bibinfo{volume}{99}:\bibinfo{pages}{191}.
\newblock \DOIprefix\doi{10.1086/131977}.
\bibitem[{{Takeda} and {Tajitsu}(2009)}]{takeda2009}
\bibinfo{author}{{Takeda}\xfnm[ Y.]}, \bibinfo{author}{{Tajitsu}\xfnm[ A.]}.
\newblock \bibinfo{title}{{High-Dispersion Spectroscopic Study of Solar Twins:
  HIP 56948, HIP 79672, and HIP 100963}}.
\newblock \bibinfo{journal}{Publications of the Astronomical Society of Japan}
  \bibinfo{year}{2009};\bibinfo{volume}{61}:\bibinfo{pages}{471}.
\newblock \DOIprefix\doi{10.1093/pasj/61.3.471}.
  \href{http://arxiv.org/abs/0901.2509}{\tt arXiv:0901.2509}.
\bibitem[{{Tody}(1986)}]{iraf1986}
\bibinfo{author}{{Tody}\xfnm[ D.]}.
\newblock \bibinfo{title}{{The IRAF Data Reduction and Analysis System}}.
\newblock In: \bibinfo{editor}{{Crawford}\xfnm[ D.L.]}, editor.
  \bibinfo{booktitle}{Instrumentation in astronomy VI}; vol.
  \bibinfo{volume}{627} of \emph{\bibinfo{series}{Society of Photo-Optical
  Instrumentation Engineers (SPIE) Conference Series}}. \bibinfo{year}{1986},
  p. \bibinfo{pages}{733}.
\newblock \DOIprefix\doi{10.1117/12.968154}.
\bibitem[{Vanderplas(2015)}]{gatspy2015}
\bibinfo{author}{Vanderplas\xfnm[ J.]}.
\newblock \bibinfo{title}{{gatspy: General tools for Astronomical Time Seriesin
  Python}}.
\newblock \bibinfo{year}{2015}.
\newblock \URLprefix \url{https://doi.org/10.5281/zenodo.14833}.
  \DOIprefix\doi{10.5281/zenodo.14833}.
\bibitem[{{VanderPlas}(2018)}]{vanderplas2018}
\bibinfo{author}{{VanderPlas}\xfnm[ J.T.]}.
\newblock \bibinfo{title}{{Understanding the Lomb-Scargle Periodogram}}.
\newblock \bibinfo{journal}{Astrophysical Journal Supplement Series}
  \bibinfo{year}{2018};\bibinfo{volume}{236}(\bibinfo{number}{1}):\bibinfo{eid}{16}.
\newblock \DOIprefix\doi{10.3847/1538-4365/aab766}.
  \href{http://arxiv.org/abs/1703.09824}{\tt arXiv:1703.09824}.
\bibitem[{{Vere{\v{s}}} et~al.(2015){Vere{\v{s}}}, {Jedicke}, {Fitzsimmons},
  {Denneau}, {Granvik}, {Bolin} et~al.}]{veres2015}
\bibinfo{author}{{Vere{\v{s}}}\xfnm[ P.]}, \bibinfo{author}{{Jedicke}\xfnm[
  R.]}, \bibinfo{author}{{Fitzsimmons}\xfnm[ A.]},
  \bibinfo{author}{{Denneau}\xfnm[ L.]}, \bibinfo{author}{{Granvik}\xfnm[ M.]},
  \bibinfo{author}{{Bolin}\xfnm[ B.]}, et~al.
\newblock \bibinfo{title}{{Absolute magnitudes and slope parameters for 250,000
  asteroids observed by Pan-STARRS PS1 - Preliminary results}}.
\newblock \bibinfo{journal}{Icarus}
  \bibinfo{year}{2015};\bibinfo{volume}{261}:\bibinfo{pages}{34--47}.
\newblock \DOIprefix\doi{10.1016/j.icarus.2015.08.007}.
  \href{http://arxiv.org/abs/1506.00762}{\tt arXiv:1506.00762}.
\bibitem[{{Vokrouhlick{\'y}} et~al.(2007){Vokrouhlick{\'y}}, {Breiter},
  {Nesvorn{\'y}} and {Bottke}}]{vokrouhlicky2007}
\bibinfo{author}{{Vokrouhlick{\'y}}\xfnm[ D.]},
  \bibinfo{author}{{Breiter}\xfnm[ S.]}, \bibinfo{author}{{Nesvorn{\'y}}\xfnm[
  D.]}, \bibinfo{author}{{Bottke}\xfnm[ W.F.]}.
\newblock \bibinfo{title}{{Generalized YORP evolution: Onset of tumbling and
  new asymptotic states}}.
\newblock \bibinfo{journal}{Icarus}
  \bibinfo{year}{2007};\bibinfo{volume}{191}(\bibinfo{number}{2}):\bibinfo{pages}{636--650}.
\newblock \DOIprefix\doi{10.1016/j.icarus.2007.06.002}.
\bibitem[{Warner(2011)}]{warner2011}
\bibinfo{author}{Warner\xfnm[ B.D.]}.
\newblock \bibinfo{title}{Save the lightcurves!}
\newblock In: \bibinfo{booktitle}{Society for Astronomical Sciences Annual
  Symposium}; vol.~\bibinfo{volume}{30}. \bibinfo{year}{2011}, p.
  \bibinfo{pages}{19--23}.
\bibitem[{{Warner} et~al.(2009){Warner}, {Harris} and {Pravec}}]{warner2009}
\bibinfo{author}{{Warner}\xfnm[ B.D.]}, \bibinfo{author}{{Harris}\xfnm[ A.W.]},
  \bibinfo{author}{{Pravec}\xfnm[ P.]}.
\newblock \bibinfo{title}{{The asteroid lightcurve database}}.
\newblock \bibinfo{journal}{Icarus}
  \bibinfo{year}{2009};\bibinfo{volume}{202}(\bibinfo{number}{1}):\bibinfo{pages}{134--146}.
\newblock \URLprefix \url{https://minplanobs.org/MPInfo/php/lcdb.php}.
  \DOIprefix\doi{10.1016/j.icarus.2009.02.003}; \bibinfo{note}{"Updated 2023
  Feb"}.
\bibitem[{{Warner} and {Stephens}(2020)}]{warnerstephens2020}
\bibinfo{author}{{Warner}\xfnm[ B.D.]}, \bibinfo{author}{{Stephens}\xfnm[
  R.D.]}.
\newblock \bibinfo{title}{{Near-Earth Asteroid Lightcurve Analysis at the
  Center for Solar System Studies: 2019 July-September}}.
\newblock \bibinfo{journal}{Minor Planet Bulletin}
  \bibinfo{year}{2020};\bibinfo{volume}{47}(\bibinfo{number}{1}):\bibinfo{pages}{23--34}.
\bibitem[{{Zappala} et~al.(1990){Zappala}, {Cellino}, {Barucci}, {Fulchignoni}
  and {Lupishko}}]{zappala1990}
\bibinfo{author}{{Zappala}\xfnm[ V.]}, \bibinfo{author}{{Cellino}\xfnm[ A.]},
  \bibinfo{author}{{Barucci}\xfnm[ A.M.]}, \bibinfo{author}{{Fulchignoni}\xfnm[
  M.]}, \bibinfo{author}{{Lupishko}\xfnm[ D.F.]}.
\newblock \bibinfo{title}{{An analysis of the amplitude-phase relationship
  among asteroids}}.
\newblock \bibinfo{journal}{A\&A}
  \bibinfo{year}{1990};\bibinfo{volume}{231}(\bibinfo{number}{2}):\bibinfo{pages}{548--560}.
\bibitem[{{Zellner} et~al.(1977){Zellner}, {Leake}, {Lebertre} and
  {Dollfus}}]{zellneretal1977}
\bibinfo{author}{{Zellner}\xfnm[ B.]}, \bibinfo{author}{{Leake}\xfnm[ M.]},
  \bibinfo{author}{{Lebertre}\xfnm[ T.]}, \bibinfo{author}{{Dollfus}\xfnm[
  A.]}.
\newblock \bibinfo{title}{{Polarimetry of Meteorites and the Asteroid Albedo
  Scale}}.
\newblock In: \bibinfo{booktitle}{Lunar and Planetary Science Conference};
  vol.~\bibinfo{volume}{8} of \emph{\bibinfo{series}{Lunar and Planetary
  Science Conference}}. \bibinfo{year}{1977}, p. \bibinfo{pages}{1041}.

\end{thebibliography}

\end{document}